\documentclass[aps,prd,reprint,nofootinbib]{revtex4-1}

\usepackage[utf8]{inputenc}
\usepackage[T1]{fontenc}
\usepackage[english]{babel}

\usepackage{amsmath, amssymb, amsthm, mathtools}
\usepackage{cancel}

\usepackage{graphicx,color}
\usepackage[dvipsnames]{xcolor}
\usepackage[caption=false]{subfig}
\usepackage{tikz}
\usepackage[compat=1.1.0]{tikz-feynman}
\tikzfeynmanset{warn luatex=false}
\usepackage[normalem]{ulem}
\usepackage{enumitem}

\theoremstyle{plain}

\theoremstyle{definition}

\theoremstyle{remark}

\newcommand{\td}{\text{d}}
\newcommand{\bphi}{\bar{\phi}}

\newcommand{\be}{\begin{eqnarray}}
\newcommand{\ee}{\end{eqnarray}}
\newcommand{\ba}{\begin{array}}
\newcommand{\ea}{\end{array}}
\newcommand{\ds}{\displaystyle}
\newcommand{\pa}[1]{\left(#1\right)}
\newcommand{\paq}[1]{\left[#1\right]}

\usepackage{hyperref}

\usepackage[scale=5,opacity=0.2,color=gray]{background}
\backgroundsetup{contents={PRD v2 draft}}

\begin{document}

\title{Matching Tidal Deformability (Wilson) Coefficients to Black Hole Love Numbers in Higher-Curvature Gravity}

\author{Luohan Wang}
\affiliation{Perimeter Institute for Theoretical Physics, Waterloo, Ontario, N2L 2Y5, Canada}
\affiliation{Department of Physics, University of Waterloo, Waterloo, Ontario, N2L 3G1, Canada}

\author{Luis Lehner}
\affiliation{Perimeter Institute for Theoretical Physics, Waterloo, Ontario, N2L 2Y5, Canada}

\author{Mait\'a Micol}
\affiliation{Instituto de Fisica Teorica, UNESP - Universidade Estadual Paulista, Sao Paulo 01140-070, SP, Brazil}
\affiliation{Department of Mathematics, King’s College London, The Strand, London WC2R 2LS, UK}

\author{Riccardo Sturani}
\affiliation{
Instituto de Fisica Teorica, UNESP - Universidade Estadual Paulista, Sao Paulo 01140-070, SP, Brazil}
\affiliation{ICTP South American Institute for Fundamental Research, Sao Paulo 01140-070, SP, Brazil}

\date{\today}

\begin{abstract}
We present a consistent mapping between tidal deformability coefficients (tidal Love numbers) and Wilson coefficients in effective field theory (EFT) descriptions of higher-curvature theories of gravity. In this work, we focus on the connection between the static response of a non-spinning black hole and the corresponding Wilson coefficient governing tidal imprints in gravitational-wave signals. We analyze a set of control cases to identify the key ingredients required for a systematic computation and matching procedure. In doing so, we highlight shortcomings in existing results that rely on the standard matching approach used in General Relativity when applied to higher-curvature gravity theories. As an explicit demonstration, we compute the relevant coefficients for cubic gravity theories. Our findings bridge an important gap in the correspondence between tidal Love numbers and Wilson coefficients in EFT extensions of General Relativity, which had not been thoroughly explored previously. 
\end{abstract}

\maketitle




\section{Introduction}

With gravitational-wave astronomy firmly established, see e.g.~\cite{GW2015,GWTC3,Abbott_2025}, and with further advances expected on short and mid-term timescales (such as through Pulsar Timing observations~\cite{Agazie_2023}, the space-based LISA mission~\cite{LISA}, and third-generation detectors~\cite{ET,Reitze:2019iox}), increasingly
stringent constraints on the theory of gravity governing observations will be achieved (e.g.~\cite{PhysRevLett.123.011102,Chia:2023tle,yunes2024gravitationalwavetestsgeneralrelativity,PhysRevD.112.084080}). To fully exploit this potential, precise theoretical predictions for gravitational waveforms from binary systems are essential. These are obtained using a range of methods, including perturbation theory, numerical relativity,
phenomenological modeling, effective field theory, and machine learning approaches, as well as hybrid schemes that combine different regimes to efficiently capture as much of the relevant physics as possible.
In particular, during the inspiral phase, perturbative and analytical techniques are powerful tools to predict expected signals as well as to provide important first-principles insights into observations. 

The effective field theory (EFT) approach to the post-Newtonian (PN) two-body problem---often referred to as PN-EFT or non-relativistic General Relativity (NRGR)---was first developed for non-spinning compact objects~\cite{goldenberger2006,goldberger2006towers}, and was later extended to incorporate spin~\cite{Porto2006Spinning,porto2007effective}.
The EFT framework has proven to be highly efficient for analytical calculations~\cite{Porto:2016pyg,Levi_2020}: it has, for example, been successfully applied to spinning binaries and extended to 4.5PN and 5PN orders~\cite{Levi_2023,Levi_2021}.
In the non-spinning case, this approach has reproduced the binding potential at 2PN~\cite{Gilmore2008}, 3PN~\cite{Foffa2011}, and 4PN~\cite{Foffa_2019jul,Foffa_2019jul1}, with 5PN \cite{Bl_mlein_2022,Porto:2024cwd,Almeida:2024lbv} and 6PN orders in progress~\cite{Brunello:2025gpf,Brunello:2026anu}. All in parallel compared and cross-checked with results obtained with the standard PN approach (see e.g.~\cite{Blanchet_2023,Blanchet:2023bwj}).

As part of the PN-EFT approach, a worldline behavior that goes beyond the leading-order point-particle geodesic approximation is included. It incorporates suitable couplings between curvature and higher-order multipole moments, thus capturing finite-size effects. These multipole moments are decomposed into intrinsic components and those induced by external fields, with the latter characterized by tidal deformability coefficients. The computation of black hole tidal Love numbers (TLNs) in General Relativity, and their matching to Wilson coefficients encoding tidal deformability in the effective action, was introduced in e.g.~\cite{Kol_2012_feb2} and explored systematically in~\cite{Hui:2020xxx,Hui:2022vbh,Ivanov:2022hlo,Ivanov2022qqt}. In GR, four-dimensional Schwarzschild black holes are known to have vanishing Love numbers, which, in turn, correspond to vanishing Wilson coefficients in the EFT description~\cite{Kol_2012_feb2,Poisson2009,DeLuca:2024ufn}. 
The latter property has been related to an enhanced symmetry condition \cite{Hui:2021vcv,Charalambous:2021kcz,Charalambous:2022rre,Parra-Martinez:2025bcu}.
  In modified gravity theories, on the
other hand, black holes may display non-zero TLNs~\cite{Cai:2019npx,Cardoso2018ptl,katagiri2024parametrizedlove} which have subtle but important
imprints on binary black hole merger waveforms. This has motivated significant efforts to compute TLNs in such theories; however, the matching between TLNs and the corresponding Wilson coefficients in the effective action has not been thoroughly explored. 
In particular, the question of whether the standard matching procedure used in General Relativity can be applied in beyond GR theories comes to the fore, as any breakdown could directly affect predictions for observables --- most notably gravitational waveforms.

In an effective field theory for gravity, certain higher-curvature contributions to the action are redundant: operators containing the Ricci tensor can be removed through a field redefinition in vacuum theories with vanishing cosmological constant. In the case where the operators are at least quadratic in the Ricci tensor, e.g. $R^2$, $R^3$ or $R_{ab}R^{ab}$ terms, the equations of motion are affected only at higher orders in the coupling, which allows one to conclude that these corrections have no impact at linear order. Other higher-derivative operators, such as $R\, R_{abcd} R^{abcd}$, linear in the Ricci tensor modify the field equations at linear order. In this case, Wilson coefficients for worldline operators in the EFT must acquire precise values such that these higher-curvature corrections have no physical impact (see appendix B of~\cite{Bernard:2025dyh}). Correctly identifying which contributions are removable and understanding their impact on the relation between Love numbers and Wilson coefficients is the central focus of this work.

To investigate this problem, we first consider a ``control'' model defined by a gravitational action featuring a higher-order interactions that can be removed via a field redefinition. In vacuum, such terms should have trivial effects~\cite{Cano2019LeadingHC,deRham:2020ejn,Bernard:2025dyh,Aly:2026otj}. Realizing this in the worldline EFT formalism requires introducing non-trivial Wilson coefficients, which can be explicitly computed~\cite{Bernard:2025dyh}. Comparing these coefficients with those obtained, e.g., from the standard matching of Love numbers to tidal coefficients in General Relativity, reveals a tension. This discrepancy can be more clearly understood with a scalar-field toy model, in which case we find that \emph{worldline Wilson coefficients are not always proportional to Love numbers} in the presence of higher-derivative operators. We further distinguish two types of contributions to the Wilson coefficient: one associated with finite-size effects and another arising from counterterms to the point-particle action. We then motivate a prescription that fixes these counterterms such that the solutions to the equations of motion from the effective action match those of the full theory and remain regular. Applying this procedure resolves the issues encountered in the control model, removing spurious contributions associated with redundant operators.

Finally, we apply this framework to two ``genuine'' beyond GR theories, which correspond to the lowest-order parity-invariant corrections to General Relativity in vacuum. (i) For a Riemann-cubic extension with interaction 
$R^{(3)} = R^{ab}{}_{cd} R^{cd}{}_{ef} R^{ef}{}_{ab}$, we compute the correct electric-type (even parity) quadrupolar Wilson coefficient $c_E^{l=2}$ and show that it is determined by finite size contributions but no point-particle counterterms. (ii) For a Riemann-cubic extension with 
interaction defined by $\tilde{R}^{(3)}=R_{abcd}R^{bgde}R^{a\,\,c}_{\,\,g\,\,e}$, we find that the analogous Wilson coefficient is determined by both
finite size and point particle contributions. Such difference stems from redundant operators linking $R^{(3)}$ and $\tilde{R}^{(3)}$.
In general, our approach provides a systematic strategy for determining Wilson coefficients in higher-curvature extensions of GR.

The structure of this work is as follows. In Section~\ref{sec:describe_RK} we describe the ``control'' cubic gravity theory and in~\ref{sec:RK_eom} we derive the vacuum equations of motion and perturbative black hole solutions. Section~\ref{sec:static_responses} computes the tidal Love numbers of a 4-dimensional spherically symmetric black hole in this theory and reveals a tension with the standard matching to Wilson coefficients. Section~\ref{sec:find_matching} demonstrates how the discrepancy can be resolved by analyzing a scalar field toy model. Section~\ref{sec:Rcubic} computes the electric-type quadrupolar Wilson coefficients $c_E^{l=2}$ for ``geniune'' Riemann-cubic extensions of General Relativity, together with the leading-order radiative contributions. We conclude in Section \ref{sec:conclusions} with some final observations and include further relevant information in the Appendices: A review of the standard definition of tidal Love numbers and their matching to Wilson coefficients~(\ref{app:defs_of_TLNs}); a detailed discussion  of counterterms for the cubic theories considered~(\ref{sec:cancelation}); 
some details on the point-particle action in General Relativity~(\ref{app:GR_pp}); 
the leading-order radiation calculation of the cubic theories studied in this work~(\ref{app:diagrams});
and the study of a quadratic theory with interaction $R_{abcd} R^{abcd}$~(\ref{sec:R2}). Throughout this paper, we use the geometric unit $G=c=1$, so that the mass and length have the same dimension. We restore $G$ in some places for convenience.

\section{Summary}\label{sec:background}

\subsection{Black hole effective action and finite size effects}

The point-particle action is the leading-order term that contributes to the worldline effective action of a general extended object. For isolated non-spinning, spherically symmetric objects (labeled by $a$) we have~\cite{goldenberger2006}
\begin{align}
    S_{\text{eff}}[x_a, g_{\mu\nu}]=& -m_a\int \text{d} \tau_a
    +  c_R^a\int \text{d} \tau_a R  \nonumber \\
    &+c_V^a\int \text{d} \tau_a R_{\mu\nu}\dot{x}_a^\mu \dot{x}_a^\nu+\cdots
\end{align}

The extra terms beyond the minimal coupling describing geodesic motion, i.e. $S_{\text{min}}=-m_a\int \text{d}\tau_a$, are organized in powers of curvature. As discussed in Ref.~\cite{goldenberger2006}, the $c_R,\, c_V$ coefficients are unphysical and can be consistently set to zero by a field redefiniton. After discarding worldline parameters proportional to Ricci tensors, we can rewrite the physical parts of the effective action in terms of the electric and magnetic components of the Weyl tensor $C_{\mu\rho\nu\sigma}$:
\begin{align}
    E_{\mu\nu}=C_{\mu\rho\nu\sigma}u^\rho u^\sigma,\quad
    B_{\mu\nu}=\frac{1}{2}\epsilon_{\mu\alpha\beta\sigma}C^{\alpha\beta}_{\,\,\,\,\nu\rho}u^\sigma u^\rho,
\end{align}
where $E_{\mu\nu},\,B_{\mu\nu}$ are traceless and orthogonal to the 4-velocity of the object $\dot x_a^\mu$. The general {(conservative)} effective action for non-spinning extended objects is then given by~\cite{goldenberger2006}
\begin{equation}
    \begin{aligned}\label{eq:BH_eff}
    S_{\text{eff}} =\, &S_{\text{min}}+c_E^a\int \text{d}\tau_a E^{\mu\nu}E_{\mu\nu} + c_B^a\int \text{d}\tau_a B^{\mu\nu}B_{\mu\nu} \\
    &+c_{E}^{a,\,l=3}\int \text{d}\tau_a  \nabla^\alpha E^{\mu\nu}\nabla_\alpha E_{\mu\nu}\\
    &+c_{B}^{a,\,l=3}\int \text{d}\tau_a \nabla^\alpha B^{\mu\nu}\nabla_\alpha B_{\mu\nu}+\cdots
    \end{aligned}
\end{equation}
where we understand the symmetric-trace-free (STF) projection of operators in Eq.~\eqref{eq:BH_eff}.

In general relativity, the Wilson coefficients $c_E^l$ and $c_B^l$ describe the tidal response of a black hole to external gravitational fields, and
are proportional to the corresponding tidal Love numbers~\cite{Henry:2019xhg,Kol:2010si,Hui:2020xxx}. In this work, we focus on the quadrupolar case; therefore, we replace $c_{E,B}^{l=2}$ with $c_{E,B}$.

Further, in our exposition we find it important to distinguish between the Love numbers commonly discussed in the general-relativistic literature and the Wilson coefficients appearing in the effective field theory (EFT) description of compact objects. In GR, Love numbers characterize the linear tidal response of an object --thus accessing its natural scale-- and are defined through the asymptotic behavior of the perturbed spacetime, relating induced multipole moments to an externally applied tidal field (through a parameter $k_l$). In the EFT framework, the same physical response is encoded in coefficients multiplying tidal operators in the worldline action after short-distance degrees of freedom have been integrated out (the parameters $c$ in \ref{eq:BH_eff}). However, these Wilson coefficients should not, in general, be identified directly with Love numbers. Individual EFT coefficients are not observables: their values depend on the choice of operator basis, field redefinitions, and the treatment of redundant operators. As a result, physically equivalent EFT descriptions may assign different values to particular Wilson coefficients, while leaving all observable quantities unchanged. Only combinations of coefficients that contribute to on-shell observables are invariant under such redefinitions. By contrast, Love numbers are defined from the asymptotic multipolar response of the spacetime. The relation between EFT Wilson coefficients and Love numbers therefore requires a matching procedure performed at the level of physical observables which is not necessarily a straightforward task as often assumed.

\subsection{Determining the coefficients}

We briefly introduce our conventions and definitions here; further details are provided in Appendix.~\ref{app:defs_of_TLNs}. In previous work, it has become customary to call any of the coefficients $k_l$, $\lambda_l$, or $c_{E,B}^l$ tidal Love numbers, as they are proportional to each other in general relativity. However, as we will show, they need not be proportional in higher curvature theories; hence, some clarification of their definitions is in order. Here we only focus on the parity-even part, as we do in this paper.

As customary, we denote the dimensionless parameters $k_l$ as the \emph{tidal Love numbers}. These can be extracted from the ratio between the ``response'' and ``source'' terms for some appropriate quantity~\cite{Poisson2009,Hui:2020xxx}. For instance, the electric tidal Love number can be found through the {spherical harmonic}
expansion of the tidal part of the effective Newtonian potential $\phi =  -m/r + \phi_{\text{tidal}}$, where $\phi \equiv -(g_{00}+1)/2$,
\begin{equation}
   \phi_{\text{tidal}}\sim \sum_{l>0, m\leq |l|}r^l\bigg(1+\,\cdots\,+2k_l^{E}\bigg(\frac{R}{r}\bigg)^{2l+1}+\,\cdots\bigg)Y^{lm},
\end{equation}
and the ellipses denote higher order terms in $1/r$ both in the external tidal field and in the induced response and $Y^{lm}$ are the standard scalar spherical harmonics.

We refer to $\lambda_l$ as the \emph{tidal coefficients}, which encode the response of a compact object to an external tidal field. 
They appear at $\mathcal{O}(\epsilon)$ in the equations of motion, where $\epsilon$ is the external field strength, and thus represent a finite-size effect. 
In general relativity, the even parity coefficients $\lambda_l^{(E)}$
are determined by the $\mathcal{O}(\epsilon)$ solution to the equation~\footnote{Note that, in general relativity, $\lambda_2^{(E)}$ is related to the gravitational deformability coefficient $\mu_2$ by $\lambda_2^{(E)}=4\pi G\,\mu_2$, in see Appendix.~\ref{app:defs_of_TLNs}.}~\cite{Kol_2012_feb2}
\begin{align}\label{eq:def_lambda}
 \frac{\delta}{\delta \phi}
\Big[
S_{\text{bulk}}
+ \frac{1}{4\pi G}\frac{\lambda_l^{E}}{2l!}
\int \td\tau\,
\big(
\partial_{a_1\cdots a_{l-2}}
E_{a_{l-1}a_l}
\big)^2
\Big]=0,
\end{align}
and thus gives the relation $c_E=\lambda_2^E/16\pi G$.

We refer to $c_{E,B}^{l}$ as the even (odd) parity \emph{Wilson coefficients}.  They are determined by varying the full effective action $S_{\text{eff}}$ with all singularities canceled by counterterms. 
As mentioned, in GR, one obtains $k_l^{E,B}\sim\lambda_l^{E}\sim c_l^{E}$. However, as we describe in this work, we find this relation to be violated in higher-curvature theories, whenever point-particle counterterms are required: $c_{E}$ would receive contributions from both the point-particle level and the finite-size level equation of motion, which we denote as $c_{E}^{pp}$ and $c_{E}^{fs}$ respectively. $c_{E}^{fs}=\lambda_2/16\pi G$ and has the same interpretation as the tidal deformability in GR, while $c_E^{pp}$ is new for higher-curvature theories. Both $c_E^{pp}$ and $c_E^{fs}$ are gauge-invariant.

\section{Introducing a control theory}\label{sec:describe_RK}

\subsection{Field redefinitions and redundant contributions}\label{sec:intro_RK}

An effective field theory of gravity generically includes higher-order curvature invariants, such as $R^2$, $R^{\mu\nu}R_{\mu\nu}$, and $R^{\mu\nu\rho\sigma}R_{\mu\nu\rho\sigma}$, among others. In the absence of additional fields, the most general action can be written as~\cite{Cano2019LeadingHC}
\begin{equation}
\label{eq:lexp}
    S_{\text{bulk}}=S_{\mathrm{EH}}+\frac{1}{16\pi G}\sum_{n\geq 2}\ell^{2n-2}S^{(2n)},
\end{equation}
where $S^{(2n)}$ is constructed from curvature invariants with $2n$ derivatives of the metric, and $\ell$ is the length scale of the new physics. {The small parameter of expansion (\ref{eq:lexp}) is $\ell^2 {\cal R}$, where ${\cal R}$ denotes the curvature scale.}

For four-derivative invariants, $S^{(4)}$ is a linear combination of $R^2,\,R_{\mu\nu}R^{\mu\nu},R^{\mu\nu\rho\sigma}R_{\mu\nu\rho\sigma},R^{\mu\nu\rho\sigma}\tilde{R}_{\mu\nu\rho\sigma}$, where $\tilde{R}_{\mu\nu\rho\sigma}$ is the (left) Hodge dual of the Riemann tensor. Among these four terms, $R^{\mu\nu\rho\sigma}\tilde{R}_{\mu\nu\rho\sigma}$ is topological and $R^{\mu\nu\rho\sigma}R_{\mu\nu\rho\sigma}$ can be expressed in terms of $R^2,\,R_{\mu\nu}R^{\mu\nu}$ using the Gauss-Bonnet combination, which is also topological in 4 dimensions. Therefore, at quadratic order, the independend curvature invariants can be chosen to be $R^2$ and $R^{\mu\nu}R_{\mu\nu}$. Two equivalent arguments can be used to show that these terms do not generate new physics in vacuum at $\mathcal{O}(\ell^2)$. First, one can perform a field redefinition of the form $g_{\mu\nu}\xrightarrow[]{}g_{\mu\nu}-\ell^2(\alpha G_{\mu\nu}+\beta g_{\mu\nu} R)$ to remove them from the action (new terms might be introduced at $\mathcal{O}(\ell^4)$). Second, the same redefinition maps (vacuum) solutions from both theories as $g_{\mu\nu} = g_{\mu\nu}^{(0)}-\ell^2(\alpha G_{\mu\nu}^{(0)}+\beta g_{\mu\nu}^{(0)} R^{(0)}) + \mathcal{O}(\ell^4)$, where $g_{\mu\nu}^{(0)}$ is a Ricci flat solution in general relativity ($G_{\mu\nu}^{(0)} = R^{(0)} = 0$);  therefore,  at order $\ell^2$ there is no correction to the metric\footnote{Equivalently, one can look at the vacuum equations of motion $G_{\mu\nu} = \ell^2 T_{\mu\nu}(R_{\gamma\delta})$, where $T_{\mu\nu}$ is an effective energy momentum tensor linearly
proportional to $R_{\mu\nu}$. Since the LHS of this equation implies that $R_{\mu\nu}$ is $\mathcal{O}(\ell^2)$, the RHS denotes a contribution at $\mathcal{O}(\ell^4)$.}. Of course, this second argument can only be used if the operators are at least quadratic in the Ricci tensor.

Now consider six-derivative invariants in $S^{(6)}$, where one can have redundant terms {\em linear} in the Ricci tensor, e.g. $R\, R^{\rho\sigma\mu\nu}R_{\rho\sigma\mu\nu}$. We refer to theory ``seemingly corrected'' by this operator as the ``$RK$'' theory, where $K$ stands for the Kretschmann scalar, $K=R^{\rho\sigma\mu\nu}R_{\rho\sigma\mu\nu}$, and we denote the coupling by $\Lambda \equiv \ell^4$. In vacuum, this theory is related to Einstein gravity by a field redefinition, hence the two descriptions are physically equivalent~\footnote{Note however that at the level of the equation of motion, eqn (\ref{eq:full_EOM_RK}), one can not straighforwardly conclude that effects should arise at most $\mathcal{O}(\ell^8)$ as in the case with corrections (at least) quadratic on the Riccci tensor.}.
The same holds in the worldline EFT: matching to a vacuum black-hole solution implies that $O(\Lambda)$ effects can only shift operator coefficients, leaving observables unchanged. As pointed out in~\cite{Bernard:2025dyh}, one can analyze this theory by treating it as distinct from general relativity, with the benefit of gaining insights into a framework whose full outcome is already known. In this work, we will be interested in the calculation of tidal coefficients for black holes and their relation to the corresponding Wilson coefficients in the control $RK$ theory.

Note that the field redefinition $g_{\mu\nu}\xrightarrow[]{}g_{\mu\nu}(1-\Lambda K)$, which removes the bulk higher-curvature term, affects the Wilson coefficients $c_E$ and $c_B$ as\footnote{Here, we used that the Kretschmann scalar is given, up to Ricci terms, by the Weyl product $C^{\mu\nu\gamma\delta}C_{\mu\nu\gamma\delta}=8(E^{\mu\nu}E_{\mu\nu}-B^{\mu\nu}B_{\mu\nu})$.}
\begin{equation}
    \begin{aligned}\label{eq:field_redefinition}
    S_{\text{eff}}=\,&\frac{1}{16\pi G}\int \td^4x \sqrt{-g}\left(R+\Lambda RK\right)-m\int \text{d}\tau \\
    &\quad +c_E\int \text{d}\tau E^{\mu\nu}E_{\mu\nu}+c_B\int \text{d}\tau B^{\mu\nu}B_{\mu\nu} \\
    &\longrightarrow\,\frac{1}{16\pi G}\int \td^4x \sqrt{-g}R-m\int \text{d}\tau \\
    &\quad+\left(c_E+4\Lambda m\right)\int \text{d}\tau E^{\mu\nu}E_{\mu\nu} \\
    &\quad+\left(c_B-4\Lambda m\right)\int \text{d}\tau B^{\mu\nu}B_{\mu\nu}+\cdots
    \end{aligned}
\end{equation}

Thus, the effect of the bulk $RK$ term is the same as the additional ``tidal terms'' in the worldline effective action. From the equivalence between vacuum $RK$ and Einstein gravity, we obtain the black hole Wilson coefficients in the control theory, $c_E=-4\Lambda m$ and $c_B=4\Lambda m$. As already noted in~\cite{Bernard:2025dyh}, these specific values are required to cancel unphysical contributions arising from the $RK$ bulk term. 
In conventional calculations, $c_E$ and $c_B$ are taken to be proportional to the electric and magnetic quadrupolar tidal Love numbers. In what follows, we describe a non-spinning black hole in the $RK$ theory and employ it to compute the tidal love numbers in section \ref{sec:static_responses} revealing a clear tension.

\subsection{Equations of motion and perturbative solution for the control RK theory}\label{sec:RK_eom}
The vacuum equations of motion for the $RK$ theory are
\begin{equation}
    \begin{aligned}\label{eq:full_EOM_RK}
    G_{\mu\nu}(1+\Lambda K)-\Lambda\nabla_{\mu}\nabla_{\nu} K+\Lambda g_{\mu\nu}\nabla^2 K& \\
    +2\Lambda RR_{\mu}^{\,\,\gamma\delta\kappa}R_{\nu\gamma\delta\kappa}
    +4\Lambda \nabla_{\gamma}\nabla_{\delta}(RR_{\mu\,\,\nu}^{\,\,\gamma\,\,\delta})&=0, 
\end{aligned}
\end{equation}

which can be simplified to
\begin{equation}\label{eq:RKEOM}
    G_{\mu\nu}+\Lambda W_{\mu\nu}=0, \quad W_{\mu\nu}=-\nabla_{\mu}\nabla_{\nu} K+g_{\mu\nu}\nabla^2 K,
\end{equation}
up to $\mathcal{O}(\Lambda)$ terms containing Ricci tensors. One can verify that $g_{\mu\nu}=g^{(0)}_{\mu\nu}(1-\Lambda K^{(0)})$ solves this equation to $\mathcal{O}(\Lambda)$, where the superscript used denotes evaluation on the Ricci-flat zeroth-oder solution.

The static and spherically symmetric solution in $4$ dimensions is given by\footnote{{We work in the "mostly plus" metric signature convention: $\eta_{\mu\nu}={\rm diag(-1,+1,+1,+1)}$}, and we set $G=c=1$ unless otherwise stated.}
\begin{equation}\label{eq:BG_metric}
    \text{d}s^2=\bigg[-f(r)\,\text{d}t^2+\frac{1}{f(r)}\text{d}r^2+r^2\text{d}\Omega^2\bigg]\bigg(1-\frac{12\Lambda r_0^2}{r^6}\bigg),
\end{equation}
where $f(r)=1-r_0/r$, with $r_0=2m$ the location of the horizon and $m$ the ADM mass of the spacetime. We can perform the coordinate transformation $\rho=r\sqrt{1-12 \Lambda r_0^2/r^6}$, which brings the metric to the form
\begin{equation}
    \label{eq:RK_metric}
    \text{d}s^2=-F(\rho)\,\text{d}t^2+\frac{1}{G(\rho)}\text{d}\rho^2+\rho^2\text{d}\Omega^2,
\end{equation}
where
\begin{subequations}
\label{eq:FG}
\begin{align}
    F(\rho)&=1-\frac{r_0}{\rho}+\frac{6\Lambda(3r_0^3-2r_0^2\rho)}{\rho^7},\\
    G(\rho)&=1-\frac{r_0}{\rho}-\frac{6\Lambda(11r_0^3-12r_0^2\rho)}{\rho^7}\,.
\end{align}
\end{subequations}

We denote by $\rho_0=r_0\sqrt{1-\frac{12\Lambda}{r_0^4}}$ the location of the horizon in the new coordinates.

\section{Love numbers of black holes in the $RK$ theory}\label{sec:static_responses}

In this section, we compute Love numbers for non-spinning black holes in the $RK$ theory and relate them to Wilson coefficients {\em using the traditional way of matching}. {Here some clarification is due, like: We compute the tidal response by solving perturbatively the modified gravity eom and compare it to the coefficient of the $\int d\tau E^2$ and $\int d\tau B^2$ terms in the EFT, emphasizing the subtle issues that appear when a "redundant term", i.e. a term vanishing (perturbatively) on the equation of motion, is present.} The definitions and conventions can be found in the Appendix.~\ref{app:defs_of_TLNs}.

\subsection{Love numbers}\label{sec:RK_TLN}
Tidal coefficients are obtained through linearized perturbations of the black hole background~\cite{Flanagan_2008,Poisson2009} 
\begin{equation}
    g_{\mu\nu}(\Lambda)=g_{\mu\nu}^{\text{BG}}(\Lambda)+\epsilon h_{\mu\nu}(\Lambda)+\mathcal{O}(\Lambda^2,\,\epsilon^2),
\end{equation}
where $g_{\mu\nu}^{\text{BG}}(\Lambda)$ is the background metric defined in Eq.~\eqref{eq:RK_metric} and we only consider perturbations up to the linear order of $\Lambda$. We decompose $h_{\mu\nu}(\Lambda)$ into even and odd-parity sectors~\cite{RW_1957}. In the Regge--Wheeler gauge, the even-parity part becomes
\begin{align}\label{eq:pert_even}
    h^{\text{even}}_{\mu\nu}=
    \scalebox{0.9}{$\begin{pmatrix}
        F(\rho)\tilde{H}_0^{lm}Y^{lm} & \tilde{H}_1^{lm}Y^{lm} &0\\
        \tilde{H}_1^{lm}Y^{lm}& G(\rho)^{-1}\tilde{H}_2^{lm}Y^{lm}&0\\
        0&0&\rho^2\tilde{A}^{lm}Y^{lm}q_{AB}\\
    \end{pmatrix}
    $},\nonumber
\end{align}
where $q_{AB}$ is the $2$-sphere metric $q_{AB} = \text{diag}(1, \sin^2\theta)$ and $A,B=(\theta,\phi)$ are angular indices. $F(\rho)$ and $G(\rho)$ are defined in Eq.~\eqref{eq:FG}, while the unknown functions $\tilde{H}^{lm}_i$ and $\tilde{A}^{lm}$ only have $\rho$ dependence. Solutions of the linearized $RK$ theory can be decomposed as (we suppress spherical harmonic indices for now),
\begin{subequations}
\label{eq:RK_lin}
\begin{align}
    \tilde{H}_0(\rho)&=H_0(\rho)+\Lambda' h_0(\rho),\\
    \tilde{H}_1(\rho)&=H_1(\rho)+\Lambda' h_1(\rho),\\
    \tilde{H}_2(\rho)&=H_2(\rho)+\Lambda' h_2(\rho),\\
    \tilde{A}_0(\rho)&=A(\rho)+\Lambda' a(\rho),
\end{align}
\end{subequations}
where $\Lambda'=\Lambda/r_0^4$ is a dimensionless parameter.
By setting $\Lambda=0$ and requiring the solution to be regular at the horizon $r_0$, we obtain the static solution of the linearized Einstein equations
\begin{align}
    H_0&=H_2=P_l^2\left(\frac{2 \rho}{r_0}-1\right), \qquad H_1=0,\label{eq:H0}\\
    A&=\frac{r_0^2}{2 (l+2) \rho (\rho-r_0)}P_{l+1}^2\bigg(\frac{2
   \rho}{r_0}-1\bigg)\nonumber\\
    &+\frac{2 (l+2) \rho^2-2 (l+3) \rho r_0+r_0^2}{2 (l+2) \rho (\rho-r_0)}P_l^2\bigg(\frac{2 \rho}{r_0}-1\bigg),\label{eq:A}
\end{align}
where we assume axial symmetry for the perturbation ($m=0$) so that the spherical harmonics $Y^{lm}(\theta,\phi)$ reduce to $P_l(\cos\theta)$. This choice does not affect the results, since the tidal coefficients are independent of $m$. We thus suppress the index $m$ henceforth.

We now substitute the ansatz in~\eqref{eq:RK_lin} into the linearized version of Eq.~\eqref{eq:RKEOM}, obtaining $h_1=0$ and finding that the functions $\{h_2,a\}$ can be expressed in terms of $\{h_0,H_0\}$. Setting $l=2$, we have
\begin{align}
    h_2&=h_0+576\frac{r_0^3}{\rho^3},\\
    a&=\frac{\rho^6 r_0 (\rho-r_0) h_0'+\rho^5 h_0 \left(4 \rho^2-2 \rho r_0-r_0^2\right)}{4 \rho^6 (\rho-r_0)}\notag\\
   &+\frac{72 r_0^3 \left(32 \rho^4-36 \rho^3 r_0+13 \rho
   r_0^3-4 r_0^4\right)}{4 \rho^6 (\rho-r_0)}.
\end{align}

The $(00)$ component of the equations of motion give
\begin{equation}
\begin{aligned}
    &\rho^7 (\rho-r_0)^2 h_0''+\rho^6 \left(2 \rho^2-3 \rho r_0+r_0^2\right) h_0' \\
    &-\rho^5\left(6 \rho^2-6 \rho
   r_0+r_0^2\right)h_0 \\
   &-72 r_0^4 \left(24 \rho^3+26 \rho^2 r_0-69 \rho r_0^2+24 r_0^3\right)=0.
\end{aligned}
\end{equation}

Requiring the function $h_0$ to be regular at the horizon $\rho_0=r_0+O(\Lambda)$, we obtain 
\begin{equation}
    h_0=-288\frac{r_0^3}{\rho^3}-144\frac{r_0^4}{\rho^4}+72\frac{r_0^5}{\rho^5},
\end{equation}

Consequently, for $l=2$, the solution for $\tilde{H}_0$ is
\begin{equation}
    \label{eq:electric_sol}
    \tilde{H}_0=-12\frac{\rho^2}{r_0^2}\Big(1-\frac{r_0}{\rho}\Big)-\Lambda'\Big(288\frac{r_0^3}{\rho^3}+144\frac{r_0^4}{\rho^4}-72\frac{r_0^5}{\rho^5}\Big),
\end{equation}
which gives the electric tidal coefficient
\begin{equation}
    k_2^E=12\Lambda',
\end{equation}
while $k_l^E=0$ for all the other values\footnote{Strictly speaking $k_{l>2}=0$ in the coordinates of Eq.~\ref{eq:BG_metric}. Notice for $l\ge3$ the $\rho^{-(l+1)}$ coefficient is coordinate dependent.} of $l$.
The definition of $k_2^E$ can be found in Eq.~(\ref{eq:TLN_electric}) --- one essentially computes the ratio of the $r^{l}$ and $r^{-(l+1)}$ terms in $\tilde{H}_0$. Using the normalization given in Refs.~\cite{Kol:2010si,Hui:2020xxx}, the corresponding electric Wilson coefficient $c_E$ results,
\begin{equation}
    c_E=2\Lambda r_0=4\Lambda m,
\end{equation}
while $c_E^l=0$, for $l> 2$.

Magnetic tidal Love number are associated with the odd-parity sector of the perturbation, which in Regge--Wheeler gauge is given by~\cite{Cai:2019npx}
\begin{align}\label{eq:pert_odd}
    h^{\text{odd}}_{\mu\nu}=
    \scalebox{1}{$\begin{pmatrix}
        0 & 0 &\tilde{u}_0^{lm}S_A^{lm}\\
         0 & 0 &\tilde{u}_1^{lm}S_A^{lm}\\
        \tilde{u}_0^{lm}(S_A^{lm})^T&\tilde{u}_1^{lm}(S_A^{lm})^T&0
    \end{pmatrix}
    $},\nonumber
\end{align}
where $S_A^{lm}=(-\partial_\phi Y^{lm}/\sin\theta,\sin\theta\, \partial_\theta Y^{lm})$ are vector spherical harmonics of odd-parity and $\tilde{u}^{lm}_i$ are again only functions of $\rho$. Similarly, we decompose the odd-parity solutions as
\begin{subequations}
    \begin{align}
    \tilde{u}_0(\rho)&= U_0(\rho)+\Lambda' u_0(\rho),\\
    \tilde{u}_1(\rho)&= U_1(\rho) +\Lambda' u_1(\rho).
    \end{align}
\end{subequations}
where $U_0$ and $U_1$ are the perturbative solutions in general relativity. In the static limit,
\begin{equation}
    {U_0=\frac{\rho^2}{r_0}\prescript{}{2}{F_1}\left[1-l,2+l,4;\frac{\rho}{r_0}\right],} \quad U_1=0,
\end{equation}
with $\prescript{}{2}{F_1}(a,b,c;z)$ the hypergeometric function. For the $RK$ theory, we obtain $\tilde u_1(\rho)=0$ and, for $l=3$, 
\begin{equation}\label{eq:magnetic_l3}
    \tilde{u}_0=r_0\bigg[\frac{3 \rho^4}{2 r_0^4}-\frac{5 \rho^3}{2 r_0^3}+\frac{\rho^2}{r_0^2}+\Lambda'\Big(18\frac{r_0^2}{\rho^2}-15\frac{r_0^3}{\rho^3}\Big)\bigg],
\end{equation}
To extract magnetic Love numbers, one computes the ratio of the $\rho^{l+1}$ and $\rho^{-l}$ terms in $\tilde{u}_0$~\cite{Poisson2009}. However, the $\rho^{-3}$ term in Eq.~(\ref{eq:magnetic_l3}) is subleading when compared to $\rho^{-2}$, and it actually comes from a mixing of the source series~\cite{Charalambous_2022}. In fact, for the magnetic Love numbers, one will find that
\begin{equation}
    k_l^B=0,
\end{equation}
for all $l$, as for $l>2$, $\tilde{u}_0$ does not contain terms with the power $\rho^{-l}$. Consequently, the usual matching to magnetic Wilson coefficients would give $c_B^l=0$.\\

Here we stress the tension. The {\em traditional} way of computing the Wilson coefficients gives the quadrupolar coefficients $c_E=4\Lambda m$ and $c_B=0$, which disagrees with the (correct) values found in Section~\ref{sec:describe_RK}, i.e. $c_E=-4\Lambda m$ and $c_B=4\Lambda m$. 

In other words, the addition of redundant terms does not allow 
an immediate identification of the Wilson coefficient.

Note that the computation of the metric perturbation in other gauges, such as the one from direct field redefinition in~\eqref{eq:BG_metric} or the isotropic gauge in the NRG decomposition~\eqref{eq:NRG_gauge}, gives the same results for $k_l^{(E,B)}$.

One could wonder, along the lines discussed in~\cite{Gralla:2017djj}, that this inconsistency comes from a mixing of the source and response series in the $RK$ theory. 
{However, as we  demonstrate later, this is not the origin of the issue.} We conclude that in a higher-order-curvature theory, the calculation of Wilson coefficients and their matching to tidal Love numbers should be revisited. We do so next, focusing on the electric case $c_E$, as it contributes to observables at leading order.

\section{Tidal coefficients and ddetermination of effective worldline terms}
\label{sec:find_matching}
\subsection{A toy model of a scalar field}\label{sec:scalar_field}

We first examine a scalar field toy model exhibiting analogous ambiguities in the calculation of Wilson coefficients. We will see that in theories with higher-order-corrections, the Wilson coefficients receive contributions not only from tidal, i.e. finite-size effects, but also from counterterms {regularizing} the point-particle action.

Consider a real massless scalar field coupled to a localized source of radius $R$, with $R\ll\lambda$ where $\lambda$ is the characteristic { length scale variation} of the field. In this regime, the static approximation applies and we write the action as
\begin{equation}
\begin{aligned}\label{eq:scalar_original}
    S_{\text{eff}}=&-\frac{1}{2}\int \text{d} ^4 x \sqrt{-g} (\nabla \phi)^2 -g \int \text{d}  \tau \phi \\ &+\sum_{l'\ge 1}\frac{\lambda_{l'}}{2l'!}\int \text{d}  \tau (\nabla_{L'} \phi)^2,
\end{aligned}
\end{equation}
where $g$ is the coupling strength of the field to the effective point particle and $\tau$ is the proper time along the center-of-mass worldline. The (``finite size'') tidal Wilson coefficients are defined as $c^{fs}_{l'}=\lambda_{l'}/(2{l'!})$. 

We use the multi-index $L'=i_1\cdots i_{l'}$, and define $\nabla_{L'}\phi\equiv\nabla_{i_1}\cdots \nabla_{i_{l'}}\phi$. {As it defines worldline operators, the multi-index $L'$ should be understood as an STF index; however, for notation simplicity, we omit the traceless condition throughout the paper.}
Assuming a flat background, we henceforth replace covariant derivatives $\nabla$ with partial derivatives $\partial$ and $g_{\mu\nu}$ with $\eta_{\mu\nu}$. In addition, we work in a Cartesian coordinate system to simplify our notation.

{A field redefinition of the form $\phi\rightarrow\phi-\Lambda (\partial_L\phi)^2$, with $l= |L|$
a fixed integer, changes the action to}
\begin{equation}
\begin{aligned}\label{eq:Lambda_phi}
    S_{\text{eff}}=&-\frac{1}{2}\int \text{d} ^4 x \sqrt{-g}\left[(\partial \phi)^2 +2\Lambda \square\phi (\partial_{L}\phi)^2\right]  \\
    &-g \int\text{d}\tau\phi+g\Lambda\int\text{d}\tau \,(\partial_L\phi)^2\\ 
    &+\sum_{l'\ge 1}\frac{\lambda_{l'}}{2l'!}\int \text{d}  \tau (\partial_{L'} \phi)^2+\cdots.
\end{aligned}
\end{equation}
where $\square=\eta^{\mu\nu}\partial_\mu\partial_\nu$. We will take Eq.~(\ref{eq:Lambda_phi}) as the definition of the $\Lambda$ ``control'' theory in the scalar field case. The relation between solutions $\phi$ of this higher-derivative theory and the original scalar field theory $\phi_0$ takes the form
\begin{equation}\label{eq:relate_sols}
\phi=\phi_{0}+\Lambda(\partial_L\phi_{0})^2+\mathcal{O}(\Lambda^2).
\end{equation}

Denoting by $c_{l'}$ the coefficient for the worldline term $(\partial_{L'}\phi)^2$ in the ``control'' theory, we expect to fix it via
\begin{equation}
    c_l = g\Lambda+\frac{\lambda_l}{2l!},
\end{equation}
while for $l'\neq l$ the Wilson coefficients should be unaffected.

To recover the correct value of $c_l$, we must keep track of two perturbative parameters: the higher-derivative coupling $\Lambda$ and the strength of the external field $\epsilon$. Note that, when we take the point particle limit $R\rightarrow 0$ we obtain $c_l\rightarrow g\Lambda$, as $\lambda_{l}\sim R^{2l+1}$~\cite{Kol_2012_feb2,Hui:2020xxx} accounts for finite-size effects only. Thus, in the ``control'' theory, the coefficient $c_l$ decomposes into a point-particle $c_l^{pp}=g\Lambda$ and a finite-size $c_l^{fs}=\lambda_l/2l!$ contribution. In what follows, we begin with an effective action containing the higher-derivative bulk interaction  in~\eqref{eq:Lambda_phi} and show how to obtain these coefficients by imposing regularity of the solutions in the full theory.

\subsubsection*{Point-Particle}

We write the point particle part of the effective action~\eqref{eq:Lambda_phi}, now with an unknown (``point particle'') coefficient $c_l^{pp}$ to be determined,
\begin{equation}
    \begin{aligned}\label{eq:Lambda_phi2}
    S_{\text{eff}}=&-\frac{1}{2}\int \text{d} ^4 x \left[(\partial \phi)^2 +2\Lambda \square\phi (\partial_L\phi)^2\right] \\
    & -g \int\text{d}\tau\phi+c_l^{pp}\int\text{d}\tau \,(\partial_L\phi)^2.
    \end{aligned}
\end{equation}

The $c_l^{pp}$ coefficient { contains terms} canceling divergences in the on-shell fields at $\mathcal{O}(\Lambda)$, which implies $c_l^{pp} \sim \Lambda$. Note that the same scaling follows from dimensional analysis. We solve the equations of motion perturbatively in $\Lambda$ by expanding
\begin{equation}\label{eq:pp_sol_scalar_field}
    \phi=\bphi+\Lambda\delta\phi,
\end{equation}
where $\bar{\phi}$
solves the $\mathcal{O}(\Lambda^0)$ {static} equation of motion without tidal coupling:
\begin{equation}
\label{eq:bar_phi}
\square\bphi=g\,\delta(x),\quad\bphi=-\frac{g}{4\pi\,r}.
\end{equation}

At $\mathcal{O}(\Lambda)$ the equation of motion becomes
\begin{equation}
    \begin{aligned}\label{eq:eom_scalar}
    \square\delta\phi -\square(\partial_L\bphi)^2-2(-1)^l\partial_L(\square\bphi\,\partial^L\bphi) = {} & \\
    -2(-1)^l\frac{c_l^{pp}}{\Lambda}\partial_L\big(\partial^L\bphi\,\delta(x)&\big),
    \end{aligned}
\end{equation}
which, after using the leading order equation~\eqref{eq:bar_phi}, simplifies to
\begin{equation}
\label{eq:dphiLambda}
    \square(\delta\phi-(\partial_L\bphi)^2)=2(-1)^l\bigg(g-\frac{c_l^{pp}}{\Lambda}\bigg)\partial_L\big(\partial^L\bphi\, \delta(x)\big).
\end{equation}

Imposing that $\phi$ is spherically symmetric and goes to zero asymptotically, we obtain the solution using the Green's function method,
\begin{equation}\label{eq:pp_matching_scalar}
\delta\phi=(\partial_L\bphi)^2+2\bigg(1-\frac{c_l^{pp}}{g\Lambda}\bigg)\left .\partial^L\bphi\right|_{r=0}\partial_L\bphi.
\end{equation}

As a result, the regularity requirement on $\delta\phi$ fixes the value of point-particle contribution $c_{l}^{pp}$ to
\begin{equation}
    c_{l}^{pp}=g\Lambda.
\end{equation}

An analogous situation in gravity is given by the last two terms of Eq.~\eqref{eq:full_EOM_RK}, which vanish at $\mathcal{O}(\Lambda)$ by the vacuum Einstein equations and thus do not contribute to the full UV solution. However, the presence of these terms will also require the introduction of worldline counterterms in the EFT. 

\subsubsection*{Finite-size effects}

We then proceed to analyze the finite size effects encoded in $c_{l}^{fs}$, the results carry over unchanged for $l'\neq l$.
We decompose $c_l^{fs}$ as 
\begin{equation}
\label{eq:delta_cl}
    c_l^{fs}=\bar{c}_l^{fs}+\Lambda\delta c_l^{fs}, \quad \text{where} \quad \bar{c}_{l}^{fs}=\lambda_l/(2l!).
\end{equation}

To compute finite size effects, the applied tidal field and induced response are treated as a perturbation to the point-particle background solution,
\begin{equation}
\label{exp_phi}
\phi=\bphi+\Lambda\delta\phi+\epsilon\phi_l',\quad \text{with}\quad \phi_l'=\phi_l+\Lambda\delta\phi_l,  
\end{equation}
controlled by the parameter $\epsilon$. Here $\bphi$ and $\delta\phi$ has the same definition in Eq.~(\ref{eq:pp_sol_scalar_field}).

We start by computing $\phi_l$ in the original free theory {($\Lambda=0$)}. Keeping terms of $\mathcal{O}(\epsilon)$ in the variation
\begin{equation}
\frac{\delta}{\delta\phi}\left[-\frac{1}{2}\int(\partial\phi)^2+\bar{c}_l^{fs}\int\td \tau (\partial_L\phi)^2\right]=0,
\end{equation}
gives the $\mathcal{O}(\epsilon)$ equation of motion
\begin{equation}\label{eq:bk_fs_scalar}
    \square\phi_l+2(-1)^l\bar{c}_l^{fs}\partial^L\phi_l\rvert_{r=0}\,\partial_L\delta(x)=0.
\end{equation}
The {static} solution of Eq.~\eqref{eq:bk_fs_scalar} {with non-trivial angular dependence} can be written as
\begin{equation}\label{eq:sol_fs}
    \phi_l=\left(r^l+\frac{\bar b_l}{r^{l+1}}\right)P_l,
\end{equation}
which is obtained by solving the full equation of motion, and imposing boundary conditions on $r=R$ to fix $\bar{b}_l$ (see Appendix.~\ref{app:defs_of_TLNs} for details). We again substitute $Y_{lm}$ with the Legendre polynomial $P_l$ for simplicity.

Inserting $\phi_l$ into Eq.~\eqref{eq:bk_fs_scalar} and 
requiring the equation of motion to be regular at the $\mathcal{O}(r^{-l-3})$ order gives the relation $\bar{c}_l^{fs}=2\pi/(l!(2l-1)!!)\bar b_l$\footnote{The response part of $\phi_l$ gives a divergent self-energy contribution in $\partial_L\phi_l|_{r=0}\sim \partial_L\, r^{-l-1}|_{r=0}$, which can be removed by { regularizing} $\bar{c}_l^{fs}$~\cite{Poisson_2021}.
}., as described in Refs.~\cite{Kol_2012_feb2,Hui_2021}

We then proceed to consider $\delta\phi_l$, the modification to the response multiplied by $\Lambda$ in definition~(\ref{exp_phi}). Substituting the expansion~\eqref{exp_phi} into the $\mathcal{O}(\epsilon\Lambda)$ equation of motion gives 
\begin{equation}
\begin{aligned}\label{eq:epsilonLambda}
    &\square\delta\phi_l-\square\big(2\partial_L\bphi\partial^L\phi_l\big)-2(-1)^l\partial_L(\square\phi_l\partial^L\bphi)=  \\
    &-2(-1)^l\big[\bar c_l^{fs}\partial_L(\partial^L\delta\phi_l\,\delta(x)) +\delta c_l^{fs}\partial_L(\partial^L\phi_l\,\delta(x))\big]\\
    &+2(-1)^l\big[\partial_L(\square\bphi\,\partial^L\phi_l)-\frac{c_l^{pp}}{\Lambda}\partial_L(\partial^L\phi_l\,\delta(x))\big].
\end{aligned}
\end{equation}

Substituting $c_l^{pp}=g\Lambda$ into Eq.~(\ref{eq:epsilonLambda}), the third term and fourth term of the RHS cancel. The third term of the LHS and the first term of RHS scale as $r^{-3l-4}$.
and can be neglected as higher order contributions~\footnote{Equivalently, one can verify that the third term of the LHS and the first term of the RHS can be canceled by the variation of higher-order counterterms $\int \td\tau \partial_{L'}\phi\partial^{L'}(\partial_L\phi)^2$, with Wilson coefficients $-2\Lambda \bar{c}_{l'}^{fs}$.}. 
As a result, the $r^{-l-3}$ order terms in 
Eq.~\eqref{eq:epsilonLambda} gives 
\begin{equation}
    \square\big(\delta\phi_l-2\partial_L\bphi\partial^L\phi_l\big)=2(-1)^{l+1}\delta c_l^{fs}\partial^L\phi_l\rvert_{r=0}\,\partial_L\delta(x)\,.
\end{equation}

Note that since $\delta\phi_l$ arises from a field redefinition, we can use Eq.~(\ref{eq:relate_sols}) and directly obtain $\delta\phi_l=2\partial_L\bphi\partial^L\phi_l\sim r^{-l-1}$. This sets $\delta c_l^{fs} = 0$, i.e. tidal Wilson coefficients do not receive corrections in the $\Lambda$-deformed theory.

We find that the Wilson coefficient $c_l$ is, as expected,
\begin{equation}  c_l=c_l^{pp}+c_l^{fs}=g\Lambda+\frac{\lambda_l}{2l!}.
\end{equation}

The explicit expression for $\delta\phi_l$ can be found using the $\mathcal{O}(\Lambda^0)$ Eq.~\eqref{eq:bar_phi} and $\mathcal{O}(\epsilon)$ Eq.~\eqref{eq:sol_fs} solutions, 
\begin{equation}
\label{eq:dphil}
    \delta\phi_l=P_l\frac{\delta b_l}{r^{l+1}}+\mathcal{O}(R^{2l+1}r^{-(3l+2)}),
\end{equation}
with $\delta b_l = -\frac{g}{2\pi}(-1)^l\,l!\,(2l-1)!!$. We can thus see that this scalar field toy model reproduces the same issues encountered in the $RK$ theory, namely if we use the matching given in e.g. Ref.~\cite{Hui:2020xxx}, we would obtain instead the incorrect expression $\Lambda\delta  c_l^{fs} = (-1)^{l+1}g\Lambda$.

Note that to resolve the issue of Wilson coefficients, we split $c_l$ into two parts. The higher-derivative bulk vertex $\square\phi(\partial^{L}\phi)^2$ acts as an effective source in the equation of motion, but also generates non-regular solutions. These must be {regularized} by adding worldline counterterms, which defines $c_l^{pp}$. The remaining piece $c_l^{fs}$ still describes the tidal deformability; however, in this case that it is no longer simply proportional to the coefficient $b_l=\bar b_l + \Lambda\delta b_l$, which is the ratio between the $r^{-l-1}$ and $r^l$ terms in the $l$-mode perturbative solution. The variation of the bulk action is modified by the higher-order vertex, which also changes the matching relation between $b_l$ to $c_l^{fs}$. We show the EFT matching of the scalar toy model in App.~\ref{app:EFT_matching}.


\subsection{Tidal deformability Wilson coefficient in the $RK$ theory}
\label{ssec:tidalRK}
We now follow the scalar-field case to address the inconsistency of the $RK$ theory discussed in Sec.~\ref{sec:static_responses}. We work in the Kaluza-Klein decomposition of the metric field~\cite{Kol:2007rx,Kol:2010si}, i.e. $g_{\mu\nu}\to(\phi, A_i, \gamma_{ij})$, and assume a static, spherically symmetric background. In this setup, $A_i=0$, and the metric takes the form
\begin{equation}\label{eq:NRG_gauge}
    \mathrm{d}s^2 = -e^{2\phi}\mathrm{d}t^2 + e^{-2\phi}\gamma_{ij}\mathrm{d}x^i \mathrm{d}x^j.
\end{equation}

Working in the body's rest frame, the point-particle effective action,
\begin{equation}
    S_{\text{eff}} = \frac{1}{16\pi G}\int \mathrm{d}^4x \sqrt{-g}\, R - m \int \mathrm{d}\tau,
\end{equation}
reduces to~\cite{Kol:2007rx,Kol_2008}
\begin{equation}
\begin{aligned}\label{eq:staticGR_NRG}
    S_{\text{eff}}
    = &\frac{1}{16\pi G}\int \mathrm{d}t\,\mathrm{d}^3x \sqrt{\gamma}
       \big[R[\gamma] - 2\gamma^{ij}(\partial_i\phi)(\partial_j\phi) \big]\\
       &- m \int \mathrm{d}t\, e^{\phi}
\end{aligned}
\end{equation}
in the static limit of the NRG decomposition.

From the weak-field assumption, we decompose
\begin{equation}
    \gamma_{ij} = \delta_{ij} + \sigma_{ij},
\end{equation}
with $\delta_{ij}$ the flat Euclidean metric in three dimensions. In the post-Newtonian (PN) regime, both $\phi$ and $\sigma_{ij}$ admit an expansions in powers of the perturbative parameter $Gm/r$. Therefore, in the gravitational case, the perturbative expansion of $r^{-n}$ order equations is controlled by the parameters $\{Gm,\Lambda,\epsilon\}$.

At leading PN order, i.e.\ $\mathcal{O}(Gm/r)$, the point-particle equation of motion for $\phi$ is
\begin{equation}\label{eq:barphi}
    \frac{1}{4\pi G}\Box \bar{\phi} = m \delta^{(3)}(x),
    \qquad
    \bar{\phi} = -\frac{Gm}{r}.
\end{equation}

The full static background solutions for $\phi$ and $\gamma_{ij}$ correspond to the isotropic form of the Schwarzschild metric. In isotropic coordinates,
\begin{equation}
\phi = \ln\!\left(\frac{1-\rho_0/r}{1+\rho_0/r}\right)
    \sim -\frac{2\rho_0}{r},
\end{equation}
\begin{equation}
\gamma_{ij} = \Big(1-\frac{\rho_0}{r}\Big)^2\Big(1+\frac{\rho_0}{r}\Big)^2 \delta_{ij}
    \sim \bigg(1 - 2\Big(\frac{\rho_0}{r}\Big)^2\bigg)\delta_{ij},
\end{equation}
where $\rho_0 = Gm/2$, and $r$ is the radial coordinate in the isotropic metric.

Notice that the leading background contribution for $\sigma_{ij}$ enters at one PN order higher than that of $\phi$, {as it appears in the dynamics in contraction with (2 powers of) the particle velocity. Since we are only interested in the leading tidal effects at 5PN order {and in static sources}, it is consistent to approximate $\gamma_{ij} \simeq \delta_{ij}$ and neglect couplings involving $\sigma_{ij}$ at this stage.

The point particle effective action of $RK$ theory in the body's rest frame is
\begin{equation}
\begin{aligned}
    S_{\text{eff}}=&\frac{1}{16\pi G}\int \sqrt{-g}(R+\Lambda RK)-m\int \td t \,e^\phi\\
     &+c_E^{pp}\int\td t\, e^\phi E^{ij}E_{ij}.
\end{aligned}
\end{equation}

We decompose again $\phi=\bphi+\Lambda\delta\phi$, with $\delta\phi$ the leading-order correction to the GR solution $\bphi$, from the higher derivative $(\partial^2\phi)^3$ vertex. The equation of motion at $\mathcal{O}(\Lambda)$ and leading PN order is~\footnote{Here we omit the $(\square\phi)^2$ terms in $E^{ij}E_{ij}$ for simplicity.}
\begin{equation}
\begin{aligned}\label{eq:RK_pp}
    \frac{1}{4\pi G}&\square\left(\delta\phi+4\partial^{ij}\bphi\partial_{ij}\bphi\right)+\frac{2}{\pi G}\partial^{ij}\left(\square\bphi\,\partial_{ij}\bphi\right) \\& =-2\frac{c_E^{pp}}{\Lambda}\partial^{ij}\left(\partial_{ij}\bphi\,\delta(x)\right).
\end{aligned}
\end{equation}

This equation, modulo multiplicative constants, is the same as Eq.~(\ref{eq:eom_scalar}) for $l=2$. Using the same argument here, requiring that $\delta\phi$ be regular implies that $c_E^{pp}=-4\Lambda m$.

The finite-size effect part $c_E^{fs}$ is obtained using the same procedure as in the scalar field case. Denote $\phi=\bar{\phi}+\Lambda\delta\phi+\epsilon(\phi_{l=2}+\Lambda\delta\phi_{l=2})$, where $\phi_{l=2}$ is the $l=2$ perturbative solution in GR, while $\delta\phi_{l=2}$ is the correction from the $RK$ term. The equation of motion at $\mathcal{O}(\epsilon\Lambda)$ is
\begin{equation}
\begin{aligned}\label{eq:epsilonLambda_RK}
    &\frac{1}{4\pi G}\big[\square\delta\phi_{l=2} 
     +\square(8\,\partial_{ij}\bphi\, \partial^{ij}\phi_{l=2}) +8\,\partial_{ij}(\square\phi_{l=2}\,\partial^{ij}\bphi)\big]= \\
    &-2\big[\bar{c}_E^{fs}\,\partial_{ij}(\partial^{ij}\delta\phi_{l=2}\, \delta(x)) + \delta c_E^{fs}\,\partial_{ij}(\partial^{ij}\phi_{l=2}\,\delta(x))\big]\\
    &-2\big[\frac{1}{\pi G}\partial_{ij}(\square\bphi\,\partial^{ij}\phi_{l=2})+\frac{c_E^{pp}}{\Lambda}\partial_{ij}(\partial^{ij}\phi_{l=2}\,\delta(x))\big],
\end{aligned}
\end{equation}
which reduces to
\begin{equation}\label{eq:RK_matching_cfs}
\begin{aligned}
    \frac{1}{4\pi G}\big[\square\delta\phi_{l=2} 
     +\square(8\partial_{ij}\bphi\, \partial^{ij}\phi_{l=2})\big]=&\\
     -2\delta c_E^{fs}\,\partial^{ij}\phi_{l=2}\rvert_{0}\partial_{ij}&\delta(x),
\end{aligned}
\end{equation}
after setting $\bar{c}_E^{fs}=0$ and $c_E^{pp}=-4m\Lambda$, and requiring Eq.~\eqref{eq:epsilonLambda_RK} to hold up to $\mathcal{O}(r^{-5})$. This is directly analogous to the scalar-field case. Given the solution $\delta\phi_{l=2}=-8\partial_{ij}\bar{\phi}\partial^{ij}\phi_{l=2}$~\footnote{One can verify that the solution for $\delta\phi_{l=2}$ obtained from the field redefinition matches the one computed in Sec.~\ref{sec:static_responses} using the RW gauge.}, we find the expected $\delta c_E^{fs}=0$. 

The procedure of matching 
{$k^{E}_{l=2}$} to $c_E^{fs}$ adopted here is 
equivalent to the diagrammatic EFT approach proposed in Refs.~\cite{Ivanov:2022hlo,Charalambous_2022}. 
In a Riemann-cubic theory, the mixing between the source and response 
series begins at $l=2$. Collecting our results, we obtain the correct quadrupolar electric Wilson coefficient
\begin{equation}
    c_E=c_E^{pp}+c_E^{fs}=-4m\Lambda.
\end{equation}
It is { straightforward} to verify that in $RK$, one also has $c_B^{fs}=0$ and the counterterm $c_B^{pp}=4\Lambda m$, which in turn gives the quadrupolar magnetic Wilson coefficient $c_B=4\Lambda m$. We now proceed to argue that such counterterms are not special to the $RK$ theory and can indeed arise in any higher-curvature theory.

We would like to stress that the point-particle term $c_E^{pp}$ is operator basis-dependent, whose value is fixed by requiring the result of EFT computations to match that of full perturbative equations. As a result, only $c_E^{fs}$ is (proportional to) the physical tidal Love number.

\section{The effective action of a Riemann cubic theory}\label{sec:Rcubic}

As shown in the last section, counterterms are chosen to be worldline operators, whose variation cancels irregular contributions from the variation of the bulk vertex. Here, we propose a general principle for finding such counterterms.
After expanding all derivatives, if the variation of the higher-curvature operator in the bulk action contains terms of the form
\begin{equation}\label{eq:ct1}
   (\text{terms with }\phi)\,\underbrace{\partial\cdots\partial}_{n}\,(\square\phi),
\end{equation}
those will give rise to a divergent contribution for the solution $\phi$. This is because, for the leading order point-particle solution, $\square\phi\sim\delta(x)$ and $\phi\sim r^{-1}$, which leads to a solution scaling as $\left.\partial_nr^{-1}\right|_{r=0}$ at the $\mathcal{O}(\Lambda)$ order using Green's function method. On the other hand, a term like
\begin{eqnarray}\label{eq:ct2}
    (\partial\cdots\partial_i\phi)(\partial\cdots\partial^i\phi)
\end{eqnarray}
that does not contain $\square\phi$ is slightly different. It gives one regular piece of solution together with an irregular one as shown in Appendix.~\ref{sec:cancelation}. Therefore, to separate the regular and irregular parts cleanly, we rewrite such terms as
\begin{equation}
\begin{aligned}\label{eq:ct3}
    (\partial&\cdots\partial_i\phi)(\partial\cdots\partial^i\phi)=\frac{1}{2}\square\left(\partial\cdots\phi\,\partial\cdots\phi\right)\\
    &-\frac{1}{2}(\partial\cdots\square\phi)(\partial\cdots\phi)-\frac{1}{2}(\partial\cdots\phi)(\partial\cdots\square\phi).
\end{aligned}
\end{equation}

While the first term of the RHS gives a pure regular solution, the second and third term of the RHS give a pure irregular solution because they contain $\square\phi$. If such a term is of the higher order, i.e., $\partial^i\phi\partial_i\phi J[\phi]$ with $J[\phi]$ containing at least one $\phi$, one has to analyse the explicit expression of $J[\phi]$. We conclude that whenever the variation of a
$(\text{Riemann})^n$ term produces a term that would give an irregular solution; this term must be cancelled by the variation of appropriate worldline operators, as explicitly realised in Eqs.~\eqref{eq:eom_scalar} and~\eqref{eq:RK_pp}. 
Moreover, if the variation of the bulk vertex contains more than one $\square\phi$, i.e.,
    $(\square\phi)^2 S$
(with $S$ any extra contribution, that need not depend on $\phi$),
then it gives rise to a counterterm as $\int \td \tau \square\phi S$. Such operators correspond to worldline couplings with $R$ and $R_{ab}$, which give purely divergent Feymann diagrams. Equivalently, such operators can be removed by a field redefinition~\cite{Goldberger:2007hy,sturanigravitational}, and can thus be neglected. We denote such counterterms as ``trivial counterterms''. We only care about the ``non-trivial counterterms'' that are constructed by the Weyl tensor. In the remainder of this paper, we apply the prescription outlined above to determine counterterms for the case of two ``genuine''\footnote{That is, with corrections that can not be removed via field redefinitions.} cubic higher-curvature theories and obtain the corresponding quadrupolar Wilson coefficients $c_E$.

Of the five different parity-even cubic invariant terms that do not involve the Ricci tensor or its derivatives, there can be at most two independent operators, due to the symmetries of the Riemann tensor~\cite{VanNieuwenhuizen1977,Bueno2019}. We can choose them to be:
\begin{equation}
\label{eq:cubic_op}
    R^{(3)} = R^{ab}_{\,\,\,\,cd}R^{cd}_{\,\,\,\,ef}R^{ef}_{\,\,\,\,ab} \quad \text{and} \quad 
    \tilde{R}^{(3)} = R_{abcd}R^{bgde}R^{a\,\,c}_{\,\,g\,\,e},
\end{equation}
which are, moreover, related by the six-dimensional Euler density associated with the Gauss–Bonnet theorem. This density, corresponding to the cubic Lovelock term, is a topological invariant in $d=6$, while for $d<6$ it vanishes identically. As a result, for vacuum spacetimes with $d\le6$, only a single independent cubic operator remains. More explicitly, in four-dimensional vacuum spacetimes, the operators in Eq.~\eqref{eq:cubic_op} satisfy the relation\footnote{Up to terms at least quadratic in the Ricci tensor, one can write this relation schematically as $\tilde{R}^{(3)} = 1/2\, R^{(3)} - 3/8\, RK$.}
\begin{equation}
\begin{aligned}\label{eq:R3relation}
    &R_{abcd}R^{bgde}R^{a\,\,c}_{\,\,g\,\,e}=\frac{1}{2}R^{ab}_{\,\,\,\,cd}R^{cd}_{\,\,\,\,ef}R^{ef}_{\,\,\,\,ab}\\
    &~~~~~~~~~~-\frac{3}{8}RR^{abcd}R_{abcd}-3R_{abcd}R^{ac}R^{bd}\\
    &~~~~~~~~~~-4R^{a}_{\,\,b}R^{b}_{\,\,c}R^{c}_{\,\,a}+\frac{9}{2}RR^{ab}R_{ab}-\frac{5}{8}R^3.
\end{aligned}
\end{equation}
To begin, we consider the $R^{(3)}$ theory, whose bulk action is given by
\begin{equation}
    S=\frac{1}{16\pi G}\int \td ^4x \sqrt{-g}\left(R+\Lambda\,R^{ab}_{\,\,\,\,cd}R^{cd}_{\,\,\,\,ef}R^{ef}_{\,\,\,\,ab}\right),
\end{equation}
which at the leading PN order and in the static limit becomes
\begin{equation}
\begin{aligned}\label{eq:I3_action}
    S=-\frac{1}{8\pi G}\int {\rm d}^4x  \big[(\partial\phi)^2+8\Lambda(\partial^i\partial_j\phi)(\partial^j\partial_k\phi)(\partial^k\partial_i\phi)& \\ -12 \Lambda(\nabla^2\phi)(\partial_i\partial_j\phi)(\partial^i\partial^j\phi)\big]&\,,
\end{aligned}
\end{equation}
where we neglected time derivatives as we are considering the static case.


The equation of motion is given by
\begin{align}\label{eq:LHS_of_cubic}
    \frac{\delta S}{\delta \phi}=\frac{1}{4\pi G}\Big\{\nabla^2\phi+12\Lambda\big[\nabla^2\phi\nabla^4\phi+(\partial_k\nabla^2\phi)(\partial^k\nabla^2\phi)\big]\Big\}.
\end{align}


Through integration by parts, one can verify that the last two terms in Eq.~\eqref{eq:LHS_of_cubic} can be canceled by the variation of the worldline counterterm (see Appendix~\ref{sec:cancelation} for details)
\begin{equation}\label{eq:cubic_pp_action}
    S_{ct} = 6\Lambda m \int \td t \left(\nabla^2\phi\right)^2-48\Lambda m^2\pi G\int\td\tau\phi\,\nabla^2\delta(x(\tau)).
\end{equation}

Both terms in Eq.~\eqref{eq:cubic_pp_action} contain $\delta(x(\tau))$ and can be neglected as they do not contribute to diagrammatic calculations. This is an example of the so-called ``trivial counterterms'' mentioned before, which give a vanishing point-particle contribution to the Wilson coefficient, i.e. $c_E^{pp}=0$ \footnote{One can verify that the coefficient $c_B^{pp}$ is also zero, which means that the $R^{(3)}$ theory does not require any counterterms at all, since the worldline operators $E^2$ and $B^2$ are the only possible counterterms for a cubic theory.}.

At $\mathcal{O}(\Lambda)$, the equation of motion in~\eqref{eq:LHS_of_cubic} reduces to
\begin{align}
\square\delta\phi=0\xrightarrow{\phi_{r=\infty}=0}\delta\phi=0,
\end{align}
which is in agreement with the full solution in Ref.~\cite{Cai:2019npx}, i.e. the $\mathcal{O}(\Lambda)$ correction to $g_{00}$ appears at order $r^{-7}$, rather than at $r^{-6}$ as in the control $RK$ case. For the finite-size effect parts of $c_E$, the matching is as in GR and so the finite size contribution to the Wilson coefficient is proportional to $k_2$. From Refs.~\cite{Cai:2019npx,Cano:2025zyk,Cano:2026hlv}  we know that $k_2^E=28\Lambda/(Gm)^4$ and $\delta\phi_{l=2} \sim 56\, Gm/r^3$, with the decomposition of $\phi=\bphi+\Lambda\delta\phi+\epsilon(\phi_{l=2}+\Lambda\delta\phi_{l=2})$. The equation of motion for $\phi$ up to the $\mathcal{O}(\epsilon\Lambda)$ order is
\begin{equation}\label{eq:R3_matching_cfs}
   \frac{1}{4\pi G}\nabla^2\delta\phi_{l=2}
    =-2\delta c_E^{fs}\partial^{ij}\phi_{l=2}\vert_0\,\partial_{ij}\delta(x),
\end{equation}
after substituting the explicit expressions, it becomes
\begin{equation}
\square\left[Y_{2m}r^2\left(1+\frac{56\Lambda m}{r^5}-\frac{1}{N}\frac{\lambda_2^{E}}{r^5}\right)\right]=0,
\end{equation}
where $\lambda_2^{E}$ is defined in Eq.~\eqref{eq:def_lambda}. 
Using $N=4\pi/3$~\cite{Kol_2012_feb2}, regularity implies $\lambda_2^E=224\pi\Lambda Gm/3$  and, consequently, $c_E^{fs}=\lambda_2^E/16\pi G =14\Lambda m/3$. 
One thus concludes that the corresponding full Wilson coefficient $c_E$ of the Riemann cubic theory is
\begin{equation}
    c_E=c_E^{pp}+c_E^{fs}=\frac{14}{3}\Lambda m.
\end{equation}
For the $R^{(3)}$ theory, the full $c_E$ results equal to the one obtained using the matching method
{described} in Ref.~\cite{Kol_2012_feb2}.
Notice, the result $c_E^{pp}=0$ is due to specific cancellations happening for the correction   $R^{ab}_{\,\,\,\,cd}R^{cd}_{\,\,\,\,ef}R^{ef}_{\,\,\,\,ab}$. 

This does not imply that $c_E^{pp}$ are zero for all the seemingly ``genuine'' theories. To illustrate this point, we proceed to the $\tilde{R}^{(3)}$ theory (\ref{eq:cubic_op}). The leading PN bulk action is
\begin{equation}
\begin{aligned}\label{eq:tidR3_action}
    S=-\frac{1}{8\pi G}\int_x\big[(\partial\phi)^2+4\Lambda(\partial^i\partial_j\phi)(\partial^j\partial_k\phi)(\partial^k\partial_i\phi)& \\ -3\Lambda(\square\phi)(\partial_i\partial_j\phi)(\partial^i\partial^j\phi)-\Lambda (\partial^2\phi)^3\big]&,
\end{aligned}
\end{equation}
whose variation gives the equation of motion
\begin{equation}
\begin{aligned}\label{eq:LHS_of_Rtid3}
    \frac{\delta S}{\delta \phi}=\frac{1}{4\pi G}\Big\{&\square\phi-\frac{3}{2}\Lambda\square(\partial^i\partial^j\phi\partial_i\partial_j\phi)\\
    &-3\Lambda (\partial_j\partial_k\square\phi)(\partial^j\partial^k\phi) \\ 
     &+3\Lambda(\square\phi)(\square^2\phi)+\frac{3}{2}\Lambda\square(\square\phi)^2\Big\}.
\end{aligned}
\end{equation}
The last three terms can be cancelled by the variation of the counterterms (see Appendix.~\ref{sec:cancelation}):
\begin{equation}
\begin{aligned}\label{eq:ct_of_tidR3}
    S_{ct}=&\frac{3}{2}\Lambda m\int \td t (\partial^i\partial^j\phi)(\partial_i\partial_j\phi)\\
    &-12\Lambda m^2\pi G\int\td t\phi\,\square\delta(x(\tau))-\frac{3}{4}\Lambda m \int \td t (\square\phi)^2.
\end{aligned}
\end{equation}

It thus follows from the first term that $c_E^{pp}=3/2\Lambda m$ for the $\tilde{R}^{(3)}$ theory, with is in agreement with Eq.~\eqref{eq:R3relation}. 

At this point we find it important to remark the following point. One may wonder why we choose to separate $c_E$ into $c_E^{pp}$ and $c_E^{fs}$ for the case of black holes, since they are of the 
same order $\sim \Lambda m$. 
One argument for this is that $c_E^{pp}$ is introduced through a counterterm to regularize irregular self-field terms in the equation of motion. While it happens to have the same order as $c_E^{fs}$, they have disparate physical meanings and should be treated differently. In other words, $c_E^{pp}$ absorbs the singular, structure-less, behavior at the origin $r=0$, while $c_E^{fs}$ captures the object's response to external fields and is obtained by requiring that the $l$ mode perturvative solution is regular at the horizon. Therefore, we have that $c_E^{fs}$ does not appear at the point-particle level equation of motion.

We find it important to mention the difference between Eq.~(\ref{eq:RK_matching_cfs}) and Eq.~(\ref{eq:R3_matching_cfs}). In both cases, we obtain a non-zero $k_2$ by solving the full equation of motion. However, in the control $RK$ theory, the matching between $k_2$ and $c_E^{fs}$ is modified by an extra source, which yields $c_E^{fs}=0$, while in the $R^{(3)}$ theory, the matching between $k_2$ and the quadrupolar Wilson coefficient is the same as in GR. In this sense, the coefficient $k_2$ in the $RK$ theory does not encode a real ``finite-size'' effect, it is generated by a source in the bulk action and can be set to zero by a field redefinition.
Note that in Section.~\ref{sec:static_responses}, we obtained $k_2$ by solving the equation of motion for the $RK$ theory; however, the same value of $k_2$ can also be obtained from the field redefinition expression $\delta\phi_{l=2}=-8\Lambda\partial_{ij}\bar{\phi}\partial^{ij}\phi_{l=2}$, which has nothing to do with a boundary condition on the horizon.
That is, it is not describing the response of the body to tidal fields. In contrast, the value of $k_2$ in $R^{(3)}$ theory is not generated by any source in the bulk action: if we simply truncate the bulk equation of motion, i.e., the LHS of Eq.~(\ref{eq:R3_matching_cfs}) to $r^{-5}$ order, it would yield the solution $\delta\phi_{l=2}=0$. Instead, the non-zero $k_2$ in this case results from solving the full equation of motion and imposing a regular behavior at the horizon.

Along these lines, it is intriguing to raise one last point about
the result $c_{E,B}^{pp}=0$ for $R^{(3)}$ theory. Among the two equivalent choices for cubic curvature corrections to GR for parity preserving cases, the focus has been mainly on $R^{(3)}$ instead of $\tilde{R}^{(3)}$  seemingly because $R^{(3)}$ it is arguably more elegant (due to structure, symmetry, and algebraic transparency). Our analysis reveals that there might be some underlying principles in the $R^{(3)}$ theory that make $c_{E,B}^{pp}$ vanish. One may wonder if quartic theories, such as $(R^{abcd}R_{abcd})^2$~\cite{Endlich_2017}, also have special combinations that would be preferable for having vanishing counterterms, which include terms such as $\int\td \tau (\nabla E)^2$ and $\int \td\tau E^3$, along with the corresponding contributions involving the magnetic components. Such a principle would be useful for identifying at each order the optimal operators to consider in the action.

\subsection*{Radiative contributions}

In this section, we proceed to show how $c_E$ affects the leading-order radiation effective action of Riemann cubic theories. Fig.~\ref{fig:Rcub_emission} shows the diagrams generating the leading-order correction to the radiation effective action by the Riemann cubic theories; Fig.~\ref{fig:cE_emission} shows the leading-order diagrams for the radiation effective action by the worldline operator $\int\td\tau E\cdot E$, which is of the same PN order as Fig.~1~\cite{Bernard:2025dyh}. Previous research~\cite {Endlich_2017,Lins_2021,Bernard:2025dyh} have proposed a way to calculate leading-order radiative diagrams of Riemann cubic and quartic theories. However, they have been missing diagrams as Fig.~\ref{fig:Rcub_emission}.(b),(c),(d) and Fig.~\ref{fig:cE_emission}.(b),(c), which are of the same order as Fig.~\ref{fig:Rcub_emission}.(a) and Fig.~\ref{fig:cE_emission}.(a). In this paper, we calculate the radiation effective action with all these diagrams taken into account, and details can be found in the Appendix.~\ref{app:diagrams}.

\begin{figure}[t]
\begin{tikzpicture}
\draw [black] (0.,0.) -- (1.2,0.);
\draw [black] (0.,1.1) -- (1.2,1.1);
\draw [black, dash pattern=on 2pt off 1pt] (0.6,0.) -- (.6,1.1);
\draw [black, decorate, decoration={snake, amplitude=0.5mm, segment length=2mm}, shorten >=1pt] (0.6,0.5) -- (1.5,0.5);
\filldraw[black] (0.6,0.55) circle (1.5pt);
\node[align=left] at (0.6,-0.25) {$a$};
\draw [black] (2.,0.) -- (3.2,0.);
\draw [black] (2.,1.1) -- (3.2,1.1);
\draw [black, dash pattern=on 2pt off 1pt] (2.6,.55) -- (2.6,1.1);
\draw [black, dash pattern=on 2pt off 1pt] (2.6,.55) -- (2.9,.0);
\draw [black, dash pattern=on 2pt off 1pt] (2.6,.55) -- (2.3,.0);
\draw [black, decorate, decoration={snake, amplitude=0.5mm, segment length=2mm}, shorten >=1pt] (2.6,0.8) -- (3.5,0.8);
\filldraw[black] (2.6,0.55) circle (1.5pt);
\node[align=left] at (2.6,-0.25) {$b$};
\draw [black] (4.,0.) -- (5.2,0.);
\draw [black] (4.,1.1) -- (5.2,1.1);
\draw [black, dash pattern=on 2pt off 1pt] (4.6,.55) -- (4.6,1.1);
\draw [black, dash pattern=on 2pt off 1pt] (4.6,.55) -- (4.9,.0);
\draw [black, dash pattern=on 2pt off 1pt] (4.6,.55) -- (4.3,.0);
\draw [black, decorate, decoration={snake, amplitude=0.5mm, segment length=2mm}, shorten >=1pt] (4.8,0.3) -- (5.7,0.3);
\filldraw[black] (4.6,0.55) circle (1.5pt);
\node[align=left] at (4.6,-0.25) {$c$};
\draw [black] (6.,0.) -- (7.2,0.);
\draw [black] (6.,1.1) -- (7.2,1.1);
\draw [black, dash pattern=on 2pt off 1pt] (6.6,.55) -- (6.6,1.1);
\draw [black, dash pattern=on 2pt off 1pt] (6.6,.55) -- (6.9,.0);
\draw [black, dash pattern=on 2pt off 1pt] (6.6,.55) -- (6.3,.0);
\draw [black, decorate, decoration={snake, amplitude=0.5mm, segment length=2mm}, shorten >=1pt] (6.6,0.6) -- (7.7,0.6);
\filldraw[black] (6.6,0.55) circle (1.5pt);
\node[align=left] at (6.6,-0.25) {$d$};
\end{tikzpicture}
\caption{Leading-order radiative emission diagrams from the bulk action at the leading order. The black dot represents the bulk vertex. The dashed line represents $\phi$, the wavy line represents the on-shell $\sigma_{ij}$, and the solid line represents the black hole worldline.}
\label{fig:Rcub_emission}
\end{figure}

For the $R^{(3)}$ theory, Fig.~\ref{fig:Rcub_emission} gives the leading-order radiation effective action from the bulk vertex
\begin{align}
    &R^{(3)}:S_{\text{eff, rad}}^{(\text{cub})}=\int \td t\,\frac{\mu}{4}\frac{\td^2}{\td t^2}\left(r^ir^j\right)\frac{\sigma_{ij}}{m_{\text{pl}}}\,,
\end{align}
where $M=(m_1+m_2),\,\mu=m_1m_2/M,\,m_{\text{pl}}^{-2}=32\pi G$. 
While for $\tilde{R}^{(3)}$ theory, the leading-order radiation effective action from the bulk vertex is
\begin{align}
    &\tilde{R}^{(3)}:S_{\text{eff, rad}}^{(\text{cub})}=\int \td t\,\frac{\mu}{4}\frac{\td^2}{\td t^2}\left(r^ir^j-\frac{36\Lambda GM \hat n^i \hat n^j}{r^3}\right)\frac{\sigma_{ij}}{m_{\text{pl}}}\,,
\end{align}
where $\hat n^i\equiv r^i/r$,
and for $RK$, the leading-order bulk effective action is
\begin{align}\label{eq:S_rad_eff_RK}
    &RK:S_{\text{eff, rad}}^{(\text{cub})}=\int \td t\,\frac{\mu}{4}\frac{\td^2}{\td t^2}\left(r^ir^j+\frac{96\Lambda GM \hat n^i \hat n^j}{r^3}\right)\frac{\sigma_{ij}}{m_{\text{pl}}}\,.
\end{align}
For the $\int\td\tau E_{ij}E^{ij}$ operator with Wilson coefficient $c_E$, the leading-order radiation effective action as from Figure~2 is 
\begin{align}\label{eq:S_eff_rad_cE}
    &c_E:S_{\text{eff, rad}}^{(EE)}=\int \td t\,\frac{1}{4}\frac{\td^2}{\td t^2}\left(\frac{12Gm_2c_E^{(1)}n^i n^j}{r^3}\right)\frac{\sigma_{ij}}{m_{\text{pl}}}\notag\\
&~~~~~~~~~~~~~~~~~~~~~~+(1\leftrightarrow 2)\,.
\end{align}
Therefore, the complete radiative is obtained by summing $S_{\text{eff, rad}}^{(\text{cub})}$ and $S_{\text{eff, rad}}^{EE}$ with $c_E=c_E^{fs}+c_E^{pp}$. In fact, substituting the values of $c_E^{\,RK}=-4m\Lambda$, $c_E^{pp,\,R^{(3)}}=0$ and $c_E^{pp,\,\tilde{R}^{(3)}}=3/2m\Lambda$ into the expression\footnote{We denote with a supre-index the theory that each Wilson coefficient corresponds to.}, one can find the net radiation effective action for $RK$ theory is zero, and $\frac{1}{2}(S_{\text{eff, rad}}^{(\text{cub}),\,R^{(3)}}+S_{\text{eff, rad}}^{c_E^{pp},\,R^{(3)}})=S_{\text{eff, rad}}^{(\text{cub}),\,\tilde{R}^{(3)}}+S_{\text{eff, rad}}^{c_E^{pp},\,\tilde{R}^{(3)}}$, as expected from Eq.~(\ref{eq:R3relation}). 

Note that although in the Riemann cubic theories, we have three new radiative diagrams Fig.~\ref{fig:Rcub_emission} (b), (c), (d), and an extra contribution from Fig.~\ref{fig:Rcub_emission} (a) due to the modified equation of motion~\cite{Lins_2021}, they end up canceling each other in the final radiation effective action, as shown in Appendix.~\ref{app:diagrams}.  The leading 5PN-order radiation of Riemann cubic theory only depends on the physical tidal parameter effects $c_E^{fs}$, consistently with the result in Ref.~\cite{cano2026gravitationalwaveformsbinarieshigherderivative}.

\begin{figure}[t]
\begin{tikzpicture}
\draw [black] (0.,0.) -- (1.2,0.);
\draw [black] (0.,1.1) -- (1.2,1.1);
\draw [black, dash pattern=on 2pt off 1pt] (0.6,0.) -- (.6,1.1);
\draw [black, decorate, decoration={snake, amplitude=0.5mm, segment length=2mm}, shorten >=1pt] (0.6,1.1) -- (1.5,1.5);
\filldraw[black] (0.6,1.1) circle (1.5pt);
\node[align=left] at (0.6,-0.25) {$a$};
\draw [black] (2.,0.) -- (3.2,0.);
\draw [black] (2.,1.1) -- (3.2,1.1);
\draw [black, dash pattern=on 2pt off 1pt] (2.6,1.1) -- (2.9,.0);
\draw [black, dash pattern=on 2pt off 1pt] (2.6,1.1) -- (2.3,.0);
\draw [black, decorate, decoration={snake, amplitude=0.5mm, segment length=2mm}, shorten >=1pt] (2.6,1.1) -- (3.5,1.5);
\filldraw[black] (2.6,1.1) circle (1.5pt);
\node[align=left] at (2.6,-0.25) {$b$};
\draw [black] (4.,0.) -- (5.2,0.);
\draw [black] (4.,1.1) -- (5.2,1.1);
\draw [black, dash pattern=on 2pt off 1pt] (4.6,1.1) -- (4.9,.0);
\draw [black, dash pattern=on 2pt off 1pt] (4.6,1.1) -- (4.3,.0);
\draw [black, decorate, decoration={snake, amplitude=0.5mm, segment length=2mm}, shorten >=1pt] (4.8,0.55) -- (5.7,0.55);
\filldraw[black] (4.6,1.1) circle (1.5pt);
\node[align=left] at (4.6,-0.25) {$c$};
\end{tikzpicture}
\caption{Leading-order radiative emission diagrams from $\int \td\tau E_{ij}E^{ij}$. The $E^2$ interaction is represented by a black dot.}
\label{fig:cE_emission}
\end{figure}

\section{Conclusions and Discussions}\label{sec:conclusions}

In this work, we examined how to obtain quadrupolar Wilson coefficients in the effective field theory of higher-curvature gravity theories. By analyzing some representative examples in detail, we showed that in such theories, the Wilson coefficients are not always proportional to the tidal Love numbers, as they are in general relativity. They instead are determined by a ``point-particle part'' and a ``finite-size part'' contributions, and their values can be computed following the approach described here. We showed that finite-size contributions are proportional to the standard Love numbers while point-particle contributions are required to ensure a suitable
regularization of the solution. For instance, this counterterm ensures that redundant operators, which can be removed via field redefinitions, do not affect the computation of observables\footnote{Incidentally, in the process we identified a few extra diagrams previously missed in the computation of raidative contributions of Rieman cubic theories~\cite{Bernard:2025dyh} which, for redundant corrections exactly cancel out.}. 

{We note that such point-particle contributions assuming the validity of the EFT aproach are, at first sight surprising. This indicates a contribution not tied to the `structure' of the object, which would
 hint of a violation of the principle of equivalence under the corrections considered. 
 Intuitively, for purely
 higher-order curvature corrections this should not be the case as such terms should be negligible at the level of a point-particle.
 Upon the closer inspection developed in this work, one understands that such point-particle contribution precisely
 cancels out contributions to the equations of motion which would otherwise yield a violation of this principle. 
Formally, the worldline operators ($E_{ij}^2, \cdots$) encode different behaviors related to the departure from a point-particle approximation. 
 The $c^{pp}$ coefficient cancels the backreaction of the field generated by the particle on the particle itself (and appears at $O(\Lambda)$ in the equation of motion for the worldline $x^{\mu}(\tau)$). Thus, it is related to the nonlinear self-interactions of the field
 and can be regarded as a self-force effect, as the perturbation to which it reacts is sourced by the particle itselt.
 Of course, the same operators incorporate tidal effects (geodesic deviation) arising from the finite size of the object and is computed with the background curvature.
 Incidentally, we noted the value of the point-particle contribution within a particular choice of curvature correction order is sensitive to redundant operators at such order. This might argue for a preference on particular expressions to define correction operators, at a given order, that yield no point-particle contribution. 
 }
 
In this work, we focused mainly on cubic curvature operators as their contribution can affect, as described here, leading order
($l=2$) tidal effects. In particular, we elucidated how Love numbers are to be mapped to the related Wilson coefficient that appears in the gravitational waveform. As discussed, higher order operators can affect the standard mapping for $l>2$ multipoles.
Beyond the relevance of our conclusions to the value of tidal Wilson coefficients and their impact on gravitational waveforms, our work is also of relevance in the study of gravitational signals in extreme-mass ratio binaries beyond GR~\cite{Roy:2025kra}. As well,
it is of relevance to exploit 
``universal'' relations~\cite{Gupta:2017vsl} to explore neutron stars through gravitational waves.
As argued here, compact objects should be described with additional couplings beyond their mass, and such
contributions can be systematically obtained following the approach outlined in this work.

\section*{Acknowledgments}
We thank Laura Bernard, Vitor Cardoso, Suvendu Giri, Jaume Gomis,
Eric Poisson, Huan Yang, Nicolas Yunes, Dami\'an Galante, and Chawakorn Maneerat for useful discussions.
LW thanks the ICTP-SAIFR and LL thanks the Department of Astronomy at Tsinghua University
and the Observatoire de Paris, Meudon for hospitality where parts of this work
were carried out.
This work was supported by the Natural Sciences and Engineering Research Council
(NSERC) of Canada. LL also thanks financial support via the Carlo Fidani Rainer Weiss Chair
at Perimeter Institute and CIFAR.  LL is also supported in part by the Simons
Foundation through Award SFI-MPS-BH-00012593-12.
MM gratefully acknowledges funding provided by the National Council for Scientific and Technological Development (CNPq) and by the Science and Technology Facilities Council (STFC). 
Research at Perimeter Institute is supported in part by the Government of
Canada through the Department of Innovation, Science and Economic Development and by the
Province of Ontario through the Ministry of Colleges and Universities.
RS acknowledge supports from FAPESP grant n. 2022/06350-2 and 2021/14335-0, and CNPq grant 309659/2025-6.

\bibliography{v2/references}

\appendix

\section{Definitions of BH tidal Love numbers and matching in General Relativity}\label{app:defs_of_TLNs}

The deformability of an extended object by an external gravitational field is characterized by its tidal Love number. First, we consider an external gravitational potential $U_{\text{ext}}$ in Newtonian gravity, generating a quadrupolar tidal field~\cite{Maggiore2007}
\begin{equation}
    \mathcal{E}_{ij}=-\partial_i\partial_j U_{\text{ext}}\rvert_{r=0}.
\end{equation}

Note that outside the source that generates it, the external field satisfies the Laplace equation $\nabla^2 \,U_{\text{ext}}=0$ so that $\mathcal{E}_{ij}$ is symmetric and traceless. Suppose we have a slowly varying external field; it will generate a perturbation $\delta\rho$ in the equilibrium position of the self-gravitating object, which modifies the (static) leading-order (signed reversed) potential of the object as
\begin{equation}
    U_{\text{self}}(x)=G\int \td^3 x'\frac{\rho(x')+\delta \rho(x')}{|x-x'|}.
\end{equation}

In the expression above $\rho(x)$ is the equilibrium background configuration, which is assumed to be spherically symmetric, and $\delta\rho$ is the deformation from the tidal force.
Outside the object, one performs a multipole expansion, taking the origin as the center of mass (COM) of the object,
\begin{equation}
    \frac{1}{|x-x'|}=\frac{1}{r}+\frac{\hat n^i}{r^2}x'_i+\frac{3\hat{n}^i\hat{n}^j-\delta^{ij}}{2r^3}x'_ix'_j+\cdots,
\end{equation}
where $r=|x|$ and $\hat{n}^i=x^i/r$. The first term gives the Newtonian potential, while the second part is $\propto\int d^3x'\,\rho(x')x'_i$ and vanishes by the definition of the center of mass frame. The third term gives the induced (traceless) quadrupole moment
\begin{equation}
    Q_{ij}=\int \td^3 x' \, \delta\rho(x')\left(x'_ix'_j-\frac{1}{3}\delta_{ij}r'^2\right).
\end{equation}

Suppose we place the object at the equilibrium point of the external field, i.e. $\partial_i U_{\text{ext}}\rvert_{r=0}=0$~\cite{Kol_2012_feb2}. The expansion of the external field around the origin has the form $U_{\text{ext}}=-\frac{1}{2}\mathcal{E}_{ij}x^ix^j+O(r^3)$, where we ignore the constant shift generated by $U_{\text{ext}}\rvert_{r=0}$. We then obtain the Newtonian potential, adding self and external contributions, up to $l=2$ terms, 
\begin{equation}
    U=\frac{Gm}{r}+\frac{3G}{2r^3}\hat{n}^i\hat{n}^j Q_{ij}+O(r^{-4})-\frac{1}{2}\mathcal{E}_{ij}x^ix^j+O(r^3).
\end{equation}

To linear order, the induced quadrupole moment is proportional to the field moment as $Q_{ij}=-\mu_2\mathcal{E}_{ij}$, where $\mu_2$ is the quadrupolar gravitational deformability. The $l=2$ dimensionless electric tidal Love number is defined as $k_2=\frac{3G\mu_2}{2R^5}$, where $R$ is the radius of the object.
In the Newtonian approximation, $g_{00}=-1+2U$, and for 
 general $l$ multipole moments, the metric takes the form
\begin{align}\label{eq:Newton_TLN}
    g_{00}=&-1+\frac{2GM}{r}\notag\\
    &~~~~~~-\sum_{l=2}^{\infty}\frac{2}{l(l-1)}\mathcal{E}_{i_1\cdots i_l}x^{i_1\cdots i_l}\left[1+2k_l\left(\frac{R}{r}\right)^{2l+1}\right],
\end{align}
where $\mathcal{E}_{i_1\cdots i_l}=-\frac{1}{(l-2)!}\partial_{i_1\cdots i_l}U_{\text{ext}}|_{r=0}$, and we denote all the $i_1\cdots i_l$ indices as $L$. 

We make here a quick detour relating the computation of tidal Love numbers $k_l$ in Newtonian gravity to the corresponding computation of the electric polarizability of a conducting sphere in electromagnetism. We start with the general solution of the Laplace equation $\nabla^2\phi=0$,
\begin{equation}
    \phi(x)=\sum_{l=0}^{\infty}\sum_{m=-l}^l(a_{lm}r^l+b_{lm}r^{-(l+1)})Y^{lm}(\theta,\varphi),
\end{equation}
where $a_{lm},\,b_{lm}$ are free parameters fixed by boundary conditions. Suppose that the conducting sphere is placed in a constant electric field $E_0\hat z$, generated by an external potential $\phi_{\text{ext}} = -E_0 r \cos\theta$. In this example, only the dipole ($l=1$) sector is relevant and we therefore write $\phi(r,\theta) = \big(a_1 r + b_1 r^{-2} \big)\cos\theta$. The boundary condition at infinity is such that $\phi \rightarrow \phi_{\text{ext}}$ as $r\rightarrow \infty$, which fixes $a_1=-E_0$. We also impose the boundary condition for a perfect conduction at the surface of the sphere, $\phi(r=R,\theta)=0$, which implies $b_1=E_0 R^3$. Thus, we have the solution
\begin{equation}\label{eq:electric_response}
    \phi(r,\theta) =  -E_0\left(r-\frac{R^3}{r^2}\right)\cos\theta,
\end{equation}
and the ratio of $r^{-2}$ and $r^1$ terms gives the electric polarizability of a conducting sphere, $\alpha\propto R^3$. Similarly, in Newtonian gravity, the Love number $k_l$ is defined by the ratio between the coefficients of the terms in the source and response, which scale as $r^l$ and $r^{-(l+1)}$, respectively. 

While there is a clear separation between the ``source'' and ``response'' in Newtonian gravity, the same is not always true in general relativity, where the effective potential $\phi=-(g_{00}+1)/2$ is given by a PN expansion at all orders~\cite{Ivanov:2022hlo},
\begin{equation}
    \begin{aligned}\label{eq:TLN_electric}
    \phi=-\frac{Gm}{r}+\sum_{l=2}^{\infty}\frac{1}{l(l-1)}\mathcal{E}_l r^l\Big[\Big(1+c_1\frac{Gm}{r}+\cdots\Big)& \\
    +2k_l^E\Big(\frac{R}{r}\Big)^{2l+1}\Big(1+b_1\frac{Gm}{r}+\cdots\Big)&\Big].\nonumber
\end{aligned}
\end{equation}

The coefficients $k^E_l$ are denoted electric tidal Love numbers, which reduce to $k_l$ in the Newtonian limit $Gm/r\ll 1$. The series defined by $c_i$ and $b_i$ are denoted source and response series, respectively. 
Notice the above expansion shows an overlap of source and response series; indeed  at $r^{2l+1}$ one has,
\begin{equation}\label{eq:mixing_term}
    {}\qquad\left(2k^E_l+c_{2l+1}\left(\frac{Gm}{R}\right)^{2l+1}\right)R^{2l+1},
\end{equation}
which could  lead to ambiguities in the definition of Love number~\cite{Ivanov:2022hlo}.
This ambiguity can be resolved, e.g., by analytically continuing the parameter $l$ to extend its domain from $\mathbb{N}$ to $\mathbb{R}$~\cite{Kol_2012_feb2}, or through a suitable EFT calculation, where one trades Love numbers for the gauge invariant tidal Wilson coefficients~\cite{Ivanov:2022hlo}.

In general relativity, the spin-$2$ perturbation allows for an odd-parity sector, which has no analogy in Newtonian gravity and corresponds to the magnetic tidal Love number $k_l^B$. 
The odd-parity components of the perturbation in the Regge–-Wheeler gauge are given by
\begin{align}
    h_{0A}=u_0(r)\,S_A^{lm},\quad h_{rA}=u_1(r)\, S_A^{lm},
\end{align}
where $A=(\theta,\phi)$ are angular indices, $S_A^{lm} = -\epsilon^{\,\,\,B}_A D_B Y_{lm}$ are odd-parity vector spherical harmonics, $D_A$ denotes the covariant derivative compatible with the $2$-sphere metric $q_{AB} = \text{diag}(1, \sin^2\theta)$ and $\epsilon_{AB} = \sqrt{q}\,\left(\begin{smallmatrix} 0 & 1 \\ -1 & 0 \end{smallmatrix}\right)$ are the corresponding Levi-Civita tensors.

Solving Einstein equations in the static limit, one finds that $u_1$ vanishes and $u_0$ is related to the Regge--Wheeler odd-parity master variable as $\varphi=r^3\partial_r\left(\frac{u_0}{r^2}\right)$~\cite{Nagar:2005ea,Damour_2009}. There are 
different conventions in the literature to denote the magnetic tidal Love number~\cite{Henry_2020,Cai:2019npx,Cardoso_2017,Bini_2012,Nagar:2005ea,Damour_2009}, here we summarize the relation between them:
\begin{equation}
\label{eq:magln}
\varphi=r^3\partial_r\left(u_0/r^2\right)\sim \sum_{l}\left(r^{l+1}+j_l\frac{R^{2l+1}}{r^l}\right)\,,
\end{equation}
which is used in Refs.~\cite{Henry_2020,Nagar:2005ea,Bini_2012,Damour_2009}. Or equivalently,
\begin{equation*}
    u_0\sim r^{l+1}\left(1+\,\cdots\,-\frac{2(l+1)}{l}k_l^B\bigg(\frac{R}{r}\bigg)^{2l+1}+\,\cdots\right),
\end{equation*}
which is used in Refs.~\cite{Poisson2009,Cai:2019npx,Cardoso_2017}. Both $j_l$ and $k_l^B$ are invariant under infinitesimal gauge transformations.

The relationship between tidal Love numbers $k_l^E,\,k_l^B(j_l)$ and Wilson coefficients $c_E^l,\,c_B^l$ in a EFT for gravity can be obtained in two ways: by introducing the polarizability coefficients $\mu_l,\sigma_l$ as the couplings of the non-minimal tidal operators (hence proportional to the Wilson coefficients) and obtaining the Love numbers through a conventional normalization, as in Ref.~\cite{Henry:2019xhg}; or by an explicit matching procedure between the EFT and the full-theory response, as in Refs.~\cite{Kol_2012_feb2,Hui_2021}. In general relativity, the two methods give the same relation, which, for the $l=2$ mode, takes the form~\footnote{For $l=2$, one has the relation $j_{2} = 12k_2^B$.}
\begin{align}\label{eq:matching_of_TLN}
    &\mu_2=\frac{2}{3G}k_2^ER^5,\quad &c_E=\frac{\mu_2}{4},\\
    &\sigma_2=\frac{1}{48G}j_2R^5,\quad &c_B=\frac{2\sigma_2}{3}.
\end{align}

\section{Counterterms of the Riemann cubic theory}\label{sec:cancelation}
In this section, we explain our approach of finding counterterms in detail, and show explicitly how to obatin the counterterms in Eq.~(\ref{eq:cubic_pp_action}) and Eq.~(\ref{eq:ct_of_tidR3}).

First, if the variation of the bulk vertex contains terms as given by Eq.~(\ref{eq:ct1}), then at $\mathcal{O}(\Lambda^1)$, its leading-order equation of motion is
\begin{equation}
    \Lambda^1:\square\delta\phi+J[\bphi]\partial_{\cdots}\square\bphi=0,
\end{equation}
which gives the solution of $\delta\phi$ by the Green's function method
\begin{align}
    \delta\phi &\sim \int\, \td^3y J[\bphi(y)]\partial_{\cdots}\square\bphi(y) G(x-y)\\\notag
    &\sim \int \td^3y \,J[\bphi(y)]\partial_{\cdots}\delta(y) G(x-y),
\end{align}
note that, as long as $J[\phi]$ contains at least one $\phi$, it would give a solution diverging as $1/0$ due to $\int_y 1/y\, \delta(y)$. Therefore, a source like Eq.~(\ref{eq:ct1}) gives a divergent solution to $\delta\phi$ and should be regularized. Note that while the Green's function method can also gives irregular solutions in higher-PN expansions within GR, this occurs when the boundary condition at $r\xrightarrow[]{}\infty$ is not asymptotically flat and the method fails~\cite{Maggiore:2007ulw}. Our case however, gives irregular solutions due to sigularities at $r=0$.

Second, there might be terms like Eq.~(\ref{eq:ct2}) that needs to be addressed. We do so by a suitable splitting in terms of regular and irregular pieces. Here we use $\partial_i\partial_j\partial_k\phi \partial^i\partial^j\partial^k\phi$ as a particular example to describe the key strategy. By the Green's function method, such a term gives the solution
\begin{align}
    \delta\phi &\sim \int\, \td^3y \partial_i\partial_j\partial_k\bphi(y) \partial^i\partial^j\partial^k\bphi(y) G(x-y)\\\notag
    &\xrightarrow[]{\text{Integration by parts}}\\\notag
    &~~~~~=-\int \td^3y \,\partial_j\partial_k\bphi(y) \partial^j\partial^k\square\bphi(y) G(x-y)\\\notag
    &~~~~~~~~~~+\frac{1}{2}\int \td^3y \,\partial_j\partial_k\bphi(y) \partial^j\partial^k\bphi(y) \square G(x-y),
\end{align}
where the first term on the RHS gives a purely divergent solution as the case yield by Eq.~(\ref{eq:ct1}) and is dealt analogously. The second term gives a purely regular solution proportional to $\partial_i\partial_j\bphi(r)\partial^i\partial^j\bphi(r)$. This thus illustrates the approach: when encountering terms like Eq,~(\ref{eq:ct2}) in the equation of motion, we write such terms in the form in Eq.~(\ref{eq:ct3}), so that the regular piece and irregular piece are separated completely.

We {next} show how to find the counterterms in Eq.~(\ref{eq:cubic_pp_action}) and Eq.~(\ref{eq:ct_of_tidR3}):

(i). For each $\Lambda(\partial_j\partial_k\nabla^2\phi)(\partial^j\partial^k\phi)\delta\phi$ in the variation,
\begin{align}\label{eq:cta}
    &\Lambda\int\td^4 x(\partial_j\partial_k\nabla^2\phi)(\partial^j\partial^k\phi)\delta\phi\\\notag
    &=\Lambda\int\td^4 x\left[\partial_j\partial_k(4\pi Gm\delta(x))\right](\partial^j\partial^k\phi)\delta\phi+\mathcal{O}(\Lambda^2)\\\notag
    &=4\pi G m\Lambda \int\td^4x \delta(x)\partial_k\partial_j(\partial^k\partial^j\phi\delta\phi)\\\notag
    &=4\pi G m\Lambda\int \td^4 x\delta(x)\left[\nabla^4\phi\delta\phi+\partial^j\partial^k\phi\partial_j\partial_k\delta\phi\nonumber \right. \nonumber \\
    &~~~~~~~~~~~~~~~~~~~~~~~~
    ~~~~~~~~~~~~~~~\left.+2\partial^j\square\phi\partial_j\delta\phi\right],\notag
\end{align}
(ii). For each $\Lambda\partial_i\square\phi\partial^i\square\phi\delta\phi$ term in the variation,
\begin{align}\label{eq:ctb}
&\Lambda\int \td^4 x\partial_i\square\phi\partial^i\square\phi\delta\phi\\\notag
    &=4\pi G\Lambda m\int \td^4x\partial_k\delta(x)\partial^k\square\phi\delta\phi\\\notag
    &=-4\pi G\Lambda m\int \td^4x\delta(x)\left[\square^2\phi\delta\phi+(\partial^k\square\phi)(\partial_k\delta\phi)\right].
\end{align}
(iii). For each $\Lambda \square\phi\square^2\phi\delta\phi$ in the variation,
\begin{align}
    &\Lambda\int \td^4 x\square\phi\square^2\phi\delta\phi\\\notag
    &=\delta\left[16\pi^2G^2m^2\Lambda\int \td\tau \square\delta(x(\tau))\phi\right].
\end{align}
Additional bulk interaction structures
can be traded for the above via integrations by parts, as
\be
\ba{rcl}
\label{eq:comment}
\ds(\partial_{ijk}\phi)^2&=&1/2\nabla^2((\partial^i\partial^j\phi)^2)-(\partial_i \partial_j \nabla^2\phi)(\partial_i \partial_j\phi)\,,\\
\ds\nabla^2(\nabla^2\phi)^2&=&\ds 2(\nabla^2\phi)(\nabla^4\phi)+2\pa{\partial_i\square\phi}^2\,.
\ea
\ee




As to the terms inside Eq.~(\ref{eq:cta}) and Eq.~(\ref{eq:ctb}), we have\\ (a) for $\partial^j\partial^k\phi\partial_j\partial_k\delta\phi$ 
\begin{align}
    &\int \td^4 x\delta(x) \partial^j\partial^k\phi\partial_j\partial_k\delta\phi
    =\frac{1}{2}\delta\left[\int \td \tau \partial^j\partial^k\phi \partial_j\partial_k\phi\right].
\end{align}
(b) for $\square^2\phi\delta\phi$
\begin{align}
    &\int\td^4 x \square^2\phi\delta\phi\delta(x)=4\pi Gm\int\td\tau (\square \delta(x)) \delta\phi\\\notag
    &=4\pi Gm\,\delta\left[\int\td\tau\, \phi(x(\tau))\square\delta(x(\tau))\right].
\end{align}
(c). for $\partial^j\square\phi\partial_j\delta\phi$
\begin{align}\label{eq:ctofRiem}
    &\int\td^4x (\partial^j\square\phi)(\partial_j\delta\phi)\delta(x)\\\notag
    &=-\int\td^4x\square\phi(\square\delta\phi)\delta(x)-\int\td^4x(\square\phi)(\partial_j\delta\phi)(\partial^j\delta(x))\\\notag
    &=-\int \td\tau(\square\phi)(\square\delta\phi)-\int\td^4x\delta(x)(\partial_j\delta\phi)(\partial^j\square\phi)\\\notag
    &\xrightarrow[]{}\int \td t (\partial^j\square\phi)(\partial_j\delta\phi)=-\frac{1}{2}\delta\left[\int \td t (\square\phi)^2\right].
\end{align}

Substituting them into Eq.~(\ref{eq:cta}) and Eq.~(\ref{eq:ctb}), one can find the corresponding counterterms respectively. Note that we have to add a minus sign in front of these expressions to obatin the counterterms, since the variation of a counterterm cancels that of the bulk vertex.

\section{Discussions over the point-particle action in GR}\label{app:GR_pp}
In this section, we discuss how to solve the on-shell metric in (static) PN expansions using the NRG field decomposition.  In the static limit of GR, the action for an isolated point-particle source reduces to the form as in Eq.~(\ref{eq:staticGR_NRG}), whose PN solution should recover the full solutions of $\phi$ and $\gamma$ as the isotropic metric
\begin{align}\label{eq:isotropic}
    \phi=\ln\left(\frac{1-\rho_0/r}{1+\rho_0/r}\right),\,\gamma_{ij}=\left(1-(\rho_0/r)^2\right)^2,\,\rho_0=Gm/2.
\end{align}
Varying Eq.~(\ref{eq:staticGR_NRG}) gives the on-shell equation of motion of $\phi$
\begin{equation}\label{eq:EOM_pp_NRG}
    \frac{\delta S}{\delta\phi}=0\xrightarrow[]{}\frac{1}{4\pi G}\partial_i\left(\sqrt{\gamma}\gamma^{ij}\partial_j\phi\right)=m\delta(x)e^\phi.
\end{equation}
We can decompose $\phi$ and $\gamma_{ij}$ in PN orders as in Eq.~(\ref{eq:PN_NRG_metric}). At 0PN order, the equation of motion of $\phi^{(1)}$ is
\begin{equation}
    \frac{1}{4\pi G}\square\phi^{(1)}=m\delta(x),
\end{equation}
and gives the Newtonian potential, while we can verify that $\sigma^{(1)}=0$. 
At 1PN order, the equation of motion of $\phi$ becomes
\begin{equation}\label{eq:phi_1PN}
    \frac{1}{4\pi G}\square\phi^{(2)}=m\delta(x)\phi^{(1)},
\end{equation}
whose RHS contains the term $\phi\delta(x)$. According to our argument in the main text, such terms are unphysical and should be regularized by suitable counterterms. Therefore, the RHS of Eq.~(\ref{eq:phi_1PN}) is set to zero, and by imposing spherical symmetry together with asymptotically trivial behavior, we obtain $\phi^{(2)}=0$. While the equation of motion of $\sigma^{(2)}$ is shown in Eq.~(\ref{eq:1PN_sigma}) and gives the solution
\begin{equation}
    \sigma^{(2)}=-\frac{1}{2}\frac{(Gm)^2}{r^2}.
\end{equation}
Notice both $\phi^{(2)}$ and $\sigma^{(2)}$ agree with the expansion of the isotropic metric.

To further strengthen the argument, we proceed to examine the 2PN order solution. At 2PN, the equation of motion of $\phi^{(2)}$ becomes
\begin{equation}
    \frac{1}{4\pi G}\square\phi^{(3)}+\frac{1}{8\pi G}\left(\partial^j\sigma^{(2)}\partial_j\phi^{(1)}\right)=\frac{1}{2}m\delta(x)\left(\phi^{(1)}\right)^2,
\end{equation}
and using the same argument, we set the RHS to zero. Substituting our solutions of $\sigma^{(2)}$ and $\phi^{(1)}$ into the LHS, we obtain
\begin{equation}
    \phi^{(3)}=-\frac{1}{12}\frac{(Gm)^3}{r^3},
\end{equation}
which also matches the expansion of $\phi$ in Eq.~(\ref{eq:isotropic}). Therefore, we claim that after regularizing all the $\phi\,\delta(x)$ terms in the source, we can recover that on-shell solution of $\phi,\gamma_{ij}$ as the isotropic metric. 
We leave the verification to interested readers.
\section{EFT matching}
\label{app:EFT_matching}

In this section we analyze the effect of the field redefinition studied in the toy model scalar case of Sec.~\ref{sec:scalar_field}, leading to the action~\eqref{eq:Lambda_phi}. 

For the EFT analysis, the field redifinition (let us consider $|L|=2$ for simplicity)  
\be
\label{eq:phi_redef}
\phi\to \phi-\Lambda \pa{\partial_i\partial_j\phi}^2
\ee
introduces two diagrams to determine
the static $\phi$ solution, at $O(\Lambda)$, see (the last two diagrams of) Fig.~\ref{fig:ppRK}.

\begin{figure}[t]
\begin{tikzpicture}
\draw [black] (0.,0.) -- (1.2,0.);
\draw [black, dash pattern=on 2pt off 1pt] (0.6,0.) -- (.6,1.1);
\node[black] at (0.6,-0.2) {$g$};
\draw [black] (2.,0.) -- (3.2,0.);
\draw [black, dash pattern=on 2pt off 1pt] (2.2,0.) -- (2.6,.55);
\draw [black, dash pattern=on 2pt off 1pt] (3.0,0.) -- (2.6,.55);
\draw [black, dash pattern=on 2pt off 1pt] (2.6,.55) -- (2.6,1.1);
\node[black] at (2.8,0.6) {$\Lambda$};
\node[black] at (2.2,-0.2) {$g$};
\node[black] at (3.,-0.2) {$g$};
\draw [black] (4.,0.) -- (5.2,0.);
\draw [black, dash pattern=on 2pt off 1pt] (4.6,0.) -- (4.6,1.1);
\draw[dashed] (4.6,0.) arc (180:0:.25);
\node[black] at (4.5,-0.2) {$c_E$};
\node[black] at (5.1,-0.2) {$g$};
\end{tikzpicture}
\caption{Diagramatic representation of the field sourced by a point particle in the RK theory: besides the contribution from the worldline $-g\int d\tau$ and $c_E\int d\tau(\partial_i\partial_j\phi)^2$ terms, one has to add the one from the bulk interaction $\Lambda\Box\phi(\partial_i\partial_j\phi)^2$.}
\label{fig:ppRK}
\end{figure}

The $\phi$ one-point function at $O(\Lambda)$ is then given by 
\begin{equation}
\label{eq:phi1pt}
\begin{array}{l}
\ds\phi(r)=-g\int\frac {d^3k}{(2\pi)^3}
e^{i\vec k\cdot \vec r}\left\{\frac 1{k^2}-g\Lambda\int\frac{d^3q}{(2\pi)^3}q^iq^j\times\right.\\
\ds\left.\paq{\frac{(q+k)^i(q+k)^j}{q^2(q+k)^2}+
2\pa{1-\frac{c_E} {g\Lambda}}\frac {k^ik^j}{k^2q^2}}\right\}\\
\ds =-\frac{g}{4\pi r}\pa{1-\frac 3{2\pi}\frac{g\Lambda}{r^5}}\,,
\end{array}
\end{equation}
The $\Lambda$ shift in the potential is consistent with its change under the field  redefinition\footnote{As the $\Lambda^0$ solution is $\bar\phi=-g/(4\pi r)$, $\partial_i\partial_j(1/r)=(-\delta_{ij}+3\hat n^i\hat n^j)/r^3\implies (\partial_i\partial_j\bar\phi)^2=3g^2/(8\pi^2r^6)$.} whereas the $(1 -\frac{c_E}{g\Lambda})$ term in (\ref{eq:phi1pt}) multiplies a divergent term localized on the world-line ($\phi$ value on the top of the source)
which vanishes in dimensional
regularization.

\begin{figure}[t]
\begin{tikzpicture}
\draw [black] (0.,0.) -- (1.2,0.);
\draw [black, dash pattern=on 2pt off 1pt] (0.6,0.) -- (.2,1.1);
\draw [black, dash pattern=on 2pt off 1pt] (0.6,0.) -- (1.,1.1);
\node[black] at (.2,1.1) {$\otimes$};
\draw [black] (2.,0.) -- (3.2,0.);
\draw [black, dash pattern=on 2pt off 1pt] (2.6,0.) -- (2.6,.55);
\draw [black, dash pattern=on 2pt off 1pt] (2.6,.55) -- (3.,1.1);
\draw [black, dash pattern=on 2pt off 1pt] (2.6,.55) -- (2.2,1.1);
\node[black] at (2.2,1.1) {$\otimes$};
\node[black] at (0.5,-0.2) {$c_E$};
\node[black] at (2.5,-0.2) {$g$};
\node[black] at (2.8,.5) {$\Lambda$};
\end{tikzpicture}
\caption{Diagramatic representation of response to the external field in the scalar analog of $RK$ theory: besides the contribution from the worldline $c_E\int d\tau(\partial_i\partial_j\phi)^2$ term one has to add the one from the $\Lambda\Box\phi(\partial_i\partial_j\phi)^2$ in the bulk. A crossed circle denotes an external field insertion.}
\label{fig:finsizeRK}
\end{figure}

The response to a static external field can be obtained
from the expansion of the Lagrangian~\eqref{eq:Lambda_phi} around the external field $r^2\mathcal{E}_{ij}\hat n_{\langle ij\rangle}$ with constant $\mathcal{E}_{ij}$,\footnote{Indices inside brackets indicate STF projection.} one has the action for the response $\varphi$ to the external perturbation $\mathcal{E}$:

\begin{eqnarray}
S_\mathcal{E}=\mathcal{E}_{\langle ij\rangle}\Big\{-2\Lambda\int d^4x
\pa{\nabla^2\varphi}\pa{\partial_i\partial_j\bar\phi}
+\pa{\nabla^2\bar\phi}\pa{\partial_i\partial_j\varphi}\nonumber\\
+2c_E\int dt\partial_i\partial_j\varphi\Big\}\,.
\end{eqnarray}

The processes in Fig.~\ref{fig:finsizeRK}, where
the cross denotes the static external field $\mathcal{E}$, give
\begin{equation}
\label{eq:vphiEFT}
\begin{array}{rcl}
\ds\varphi(x)&=&\ds 2{\mathcal E}_{\langle ij\rangle}\pa{2g\Lambda-c_E}\int \frac{d^3k}{(2\pi)^3} e^{i\vec k\cdot \vec x}\frac{k^ik^j}{k^2}\\
 &=&\ds \frac 3{2\pi}\pa{c_E-2g\Lambda}\mathcal{E}_{\langle ij\rangle}\frac{\hat n^i\hat n^j}{r^3}\,.
 \end{array}
 \end{equation}
 
The field redefinition \eqref{eq:phi_redef} has the effect of transforming the original quadrupolar perturbation, where we assume that $c_E=0$ in the free theory EFT, as
 \be
r^2\mathcal E_{ij}\hat n_{\langle ij\rangle}\to 
  r^2\mathcal E_{ij}\hat n_{\langle ij\rangle}\pa{1-\frac{3\Lambda g}{2\pi r^5}}+O(g^2)\,.
\ee
  This is consistent with eq.~(\ref{eq:vphiEFT}) if one fixes $c_E=g\Lambda$, the same shift to the quadrupolar Wilson coefficient that appears in~\eqref{eq:Lambda_phi}.

\section{Diagrams for the leading-order radiation effective action}\label{app:diagrams}
In this section, we show the explicit calculations of Figure.~\ref{fig:Rcub_emission} and Figure.~\ref{fig:cE_emission}. The results of Figure.~\ref{fig:Rcub_emission} (a) and Figure.~\ref{fig:cE_emission} (a) have been obtained using the methods provided in Refs.~\cite{Bernard:2025dyh,Lins_2021}. For Figure.~\ref{fig:Rcub_emission} (a), the corresponding effective actions for Riemann cubic theories are
\begin{align}
    &R\,K:S_{\text{eff, rad}}^{(\text{cub})}=\int \td t\,\frac{\mu}{4}\frac{\td^2}{\td t^2}\left(r^ir^j+\frac{96\Lambda GM n^i n^j}{r^3}\right)\frac{\sigma_{ij}}{m_{\text{pl}}}\notag\\
    &~~~~~~~~~~~~~~~+72\Lambda\int \td t\,\frac{G^2M^2\mu}{r^6}n^in^j\frac{\sigma_{ij}}{m_{\text{pl}}},\label{eq:RK_fig1_a}\\
    &R^{(3)}:S_{\text{eff, rad}}^{(\text{cub})}=\int \td t\,\frac{\mu}{4}\frac{\td^2}{\td t^2}\left(r^ir^j\right)\frac{\sigma_{ij}}{m_{\text{pl}}},\\
    &\tilde{R}^{(3)}:S_{\text{eff, rad}}^{(\text{cub})}=\int \td t\,\frac{\mu}{4}\frac{\td^2}{\td t^2}\left(r^ir^j-\frac{36\Lambda GM \hat n^i \hat n^j}{r^3}\right)\frac{\sigma_{ij}}{m_{\text{pl}}}\notag\\
    &~~~~~~~~~~~~~~~-27\Lambda\int \td t\,\frac{G^2M^2\mu}{r^6}\hat n^i\hat n^j\frac{\sigma_{ij}}{m_{\text{pl}}},
\end{align}
where Figure.~\ref{fig:cE_emission}(a) gives the radiation effective action as folows
\begin{align}\label{eq:c_Efig2_a}
    &c_E:S_{\text{eff, rad}}^{(EE)}=\int \td t\,\frac{1}{4}\frac{\td^2}{\td t^2}\left(\frac{12Gm_2c_E^{(1)}\hat n^i \hat n^j}{r^3}\right)\frac{\sigma_{ij}}{m_{\text{pl}}}\notag\\
    &~~~~~~~~~~~~~~~~~~~~~~~~~+\int\td t\frac{18 G^2m_2^2c_E^{(1)}}{r^6}\hat n^i\hat n^j\frac{\sigma_{ij}}{m_{\text{pl}}}+(1\leftrightarrow 2)\,,
\end{align}
where the term in the second line comes from the mismatch between
the acceleration and its Newtonian value \cite{Lins_2021,Bernard:2025dyh}.

\subsection{Additional diagrams in Figure.~\ref{fig:Rcub_emission}}
As to Figure.~\ref{fig:Rcub_emission} (b), (c), for Riemann cubic theories, there are two types of $(\partial^2\phi)^3$ vertex: (i). $\nabla^2\phi\partial_i\partial_j\phi\partial^i\partial^j\phi$ and (ii). $\partial^i\partial_j\phi\partial^j\partial_k\phi\partial^k\partial_i\phi$. For each vertex (i), the corresponding effective Lagrangian is
\begin{align}\label{eq:Leffphi3_1}
    &\text{Fig.1 (b)+Fig.1 (c), }\nabla^2\phi\partial_i\partial_j\phi\partial^i\partial^j\phi:\notag\\\notag
    &i\mathcal{L}_{\text{eff}}^{\text{rad}}=+i\frac{m_1m_2^2}{256m^5_{\text{pl}}}\sigma_{ab}\left[\int_{\mathbf{q}}\frac{e^{i\mathbf{q}\cdot\mathbf{r}}}{q^2}q^a q^b\int_{\mathbf{k}}\frac{k^ik^j(k+q)_i(k+q)_j}{k^2(k+q)^2}\right.\\\notag
    &~~~~~~~~~~~\left. +2\int_{\mathbf{q}}\frac{e^{i\mathbf{q}\cdot\mathbf{r}}}{q^2}q^iq^j\int_{\mathbf{k}}\frac{e^{i\mathbf{k}\cdot\mathbf{r}}}{k^4}k^ak^bk_ik_j\right.\\\notag
    &~~~~~~~~~~~\left.+2\int_{\mathbf{q}}\frac{e^{i\mathbf{q}\cdot\mathbf{r}}}{q^2}q^i q^j\int_{\mathbf{k}}\frac{k^ak^b(k+q)_i(k+q)_j}{k^2(k+q)^2}
    \right]+(m_1\leftrightarrow m_2)\\
    &~~~~~~~~~=i\frac{9}{4}\frac{G^2M^2\mu}{r^6}\frac{\sigma_{ab}}{m_{\text{pl}}}n^an^b\,,
\end{align}
where $M=m_1+m_2$ and $\mu=m_1m_2/M$ and $\int_{\mathbf{k}}=\int \td^3 k/(2\pi)^3$. Here $a,\,b$ are spatial indices and we discard $\sigma_{ab}\delta^{ab}$ terms using the on-shell condition for $\sigma_{ab}$: $\sigma_{ab}\delta^{ab}=0,\,(\partial_t^2-\nabla^2)\,\sigma_{ab}=0$.

For each vertex (ii), the corresponding effective Lagrangian is 
\begin{align}\label{eq:eq:Leffphi3_2}
    &\text{Fig.1 (b)+Fig.1 (c), }\partial^i\partial_j\phi\partial^j\partial_k\phi\partial^k\partial_i\phi:\notag\\\notag
    &i\mathcal{L}_{\text{eff}}^{\text{rad}}=+i\frac{3m_1m_2^2}{256m^5_{\text{pl}}}\sigma_{ab}\left[\int_{\mathbf{q}}\frac{e^{i\mathbf{q}\cdot\mathbf{r}}}{q^4}q^a q^bq_iq_l\int_{\mathbf{k}}\frac{k^lk^j(k+q)_j(k+q)^i}{k^2(k+q)^2}\right.\\\notag    &~~~~\left.+2\int_{\mathbf{q}}\frac{e^{i\mathbf{q}\cdot\mathbf{r}}}{q^2}q_i q_l\int_{\mathbf{k}}\frac{k^ak^bk^ik^j(k+q)^l(k+q)_j}{k^4(k+q)^2}
    \right]+(m_1\leftrightarrow m_2)\\
    &~~~~=i\frac{27}{8}\frac{G^2M^2\mu}{r^6}\frac{\sigma_{ab}}{m_{\text{pl}}}\hat n^a\hat n^b\,.
\end{align}

As to Figure.~\ref{fig:Rcub_emission} (d), one needs the $(\partial^2\phi)^3\sigma$ vertices of Riemann cubic theories first, which are

\begin{widetext}
\begin{align}
\label{eq:LWRt3phi3sigma}
    &R_{abcd}R^{bgde}R^{a\,\,c}_{\,\,g\,\,e}=24\sigma^{cd}\partial_c\partial_a\phi\partial_b\partial_d\phi\partial^a\partial^b\phi-12\sigma^{cd}\nabla^2\phi(\partial_c\partial^b\phi)(\partial_b\partial_d\phi)-6\sigma^{cd}\partial_c\partial_d\phi(\partial^a\partial^b\phi)(\partial_a\partial_b\phi)\\
    \label{eq:I3phi3sigma}
    &R^{ab}_{cd}R^{cd}_{ef}R^{ef}_{ab}=48\sigma^{cd}\partial_c\partial_a\phi\partial_b\partial_d\phi\partial^a\partial^b\phi-48\sigma^{cd}\nabla^2\phi(\partial_c\partial^b\phi)(\partial_b\partial_d\phi)-24\sigma^{cd}\partial_c\partial_d\phi(\partial^a\partial^b\phi)(\partial_a\partial_b\phi)\\
    \label{eq:LWRKphi3sigma}
    &RK=-32\sigma^{cd}\nabla^2\phi(\partial_c\partial^b\phi)(\partial_b\partial_d\phi)-16\sigma^{cd}\partial_c\partial_d\phi(\partial^a\partial^b\phi)(\partial_a\partial_b\phi)\,,
\end{align}
\end{widetext}
where $a,\,b,\,c,\,d$ are all spatial indices.
Therefore, we have three types of $(\partial^2\phi)^3\sigma$ vertices:
\begin{enumerate}[label=\Roman*.]
\item $\sigma^{cd}\partial_c\partial_a\phi\partial_b\partial_d\phi\partial^a\partial^b\phi$;
\item $\sigma^{cd}\nabla^2\phi(\partial_c\partial^b\phi)(\partial_b\partial_d\phi)$;
\item $\sigma^{cd}\partial_c\partial_d\phi(\partial^a\partial^b\phi)(\partial_a\partial_b\phi)$. 
\end{enumerate}
For type (I) vertex, 
 Fig.\ref{fig:Rcub_emission}(d): $\sigma^{cd}\partial_c\partial_a\phi\partial_b\partial_d\phi\partial^a\partial^b\phi$: 
\begin{align}
\label{eq:Leffphi3sigma_1}
\notag
    &i\mathcal{L}_{\text{I}}^{\text{rad}}=+i\frac{m_1m_2^2}{256m^5_{\text{pl}}}\sigma_{ab}\left[\int_{\mathbf{q}}\frac{e^{i\mathbf{q}\cdot\mathbf{r}}}{q^2}q^i q^j\int_{\mathbf{k}}\frac{k_ik_b(k+q)_j(k+q)_a}{k^2(k+q)^2}\right.\\\notag    &~~~~\left.+2\int_{\mathbf{q}}\frac{e^{i\mathbf{q}\cdot\mathbf{r}}}{q^2}q^a q^i\int_{\mathbf{k}}\frac{k^jk^b(k+q)_j(k+q)_i}{k^2(k+q)^2}
    \right]+(m_1\leftrightarrow m_2)\\
    &~~~~~=i\frac{27}8\frac{G^2M^2\mu}{r^6}\frac{\sigma_{ab}}{m_{\text{pl}}}\hat n^a\hat n^b\,.
\end{align}

For type (II), Fig.\ref{fig:Rcub_emission}(d), $
\sigma^{cd}\nabla^2\phi(\partial_c\partial^b\phi)(\partial_b\partial_d\phi)$:
\begin{align}\label{eq:Leffphi3sigma_2}\notag
    &i\mathcal{L}_{\text{II}}^{\text{rad}}=+i\frac{m_1m_2^2}{256m^5_{\text{pl}}}\sigma_{ab}\left[\int_{\mathbf{q}}\frac{e^{i\mathbf{q}\cdot\mathbf{r}}}{q^2}q^a q_i\int_{\mathbf{k}}\frac{e^{i\mathbf{k}\cdot\mathbf{r}}}{k^2}k^b k^i
    \right]\\\notag
    &~~~~~~~~~~~~~~~~~~~~~+(m_1\leftrightarrow m_2)\\
    &~~~~~~~~~=i\frac{3}{4}\frac{G^2M^2\mu}{r^6}\frac{\sigma_{ab}}{m_{\text{pl}}}\hat n^a\hat n^b\,.
\end{align}
For type (III), Fig.\ref{fig:Rcub_emission}(d), $\sigma^{cd}\partial_c\partial_d\phi(\partial^a\partial^b\phi)(\partial_a\partial_b\phi)$:
\begin{align}\label{eq:Leffphi3sigma_3}
    \notag &i\mathcal{L}_{\text{III}}^{\text{rad}}=+i\frac{m_1m_2^2}{256m^5_{\text{pl}}}\sigma_{ab}\left[\int_{\mathbf{q}}\frac{e^{i\mathbf{q}\cdot\mathbf{r}}}{q^2}q^a q^b\int_{\mathbf{k}}\frac{k_ik_j(k+q)^i(k+q)^j}{k^2(k+q)^2}\right.\\\notag    &~~~~\left.+2\int_{\mathbf{q}}\frac{e^{i\mathbf{q}\cdot\mathbf{r}}}{q^2}q^i q^j\int_{\mathbf{k}}\frac{k_ik_j(k+q)_a(k+q)_b}{k^2(k+q)^2}
    \right]+(m_1\leftrightarrow m_2)\\
    &~~~~~=i\frac{21}{4}\frac{G^2M^2\mu}{r^6}\frac{\sigma_{ab}}{m_{\text{pl}}}n^an^b\,.
\end{align}
Therefore, the Figure.~\ref{fig:Rcub_emission} (b), (c), (d) of Riemann cubic theories are obtained by the linear combinations of Eq.~(\ref{eq:Leffphi3_1})- Eq.~(\ref{eq:Leffphi3sigma_3}). For example, Figure.~\ref{fig:Rcub_emission} (b), (c), (d) for $RK$ theory is
\begin{align}
    iL_{\text{RK}}^{\text{rad}}&=16\Lambda\,\text{Eq.~(\ref{eq:Leffphi3_1})}-32\Lambda\,\text{Eq.~(\ref{eq:Leffphi3sigma_2})}-16\Lambda\,\text{Eq.~(\ref{eq:Leffphi3sigma_3})}\notag\\
    &=-72i\Lambda\frac{G^2M^2\mu}{r^6}\frac{\sigma_{ab}}{m_{\text{pl}}}n^a n^b\,,
\end{align}
which, together with Eq.~(\ref{eq:RK_fig1_a}), gives the expression in Eq.~(\ref{eq:S_rad_eff_RK}). The full bulk radiation effective action of $R^{(3)}$, $\tilde{R}^{(3)}$ can be obtained in the same way.
The result for $R^{(3)}$ is vanishing, the one for $\tilde R^{(3)}$ can be obtained by using relation (\ref{eq:R3relation}) in which eom$^2$ are neglected.

\subsection{Diagrams in Figure.~\ref{fig:cE_emission}}

Then we proceed to calculate the effective Lagragian from Figure.~\ref{fig:cE_emission} (b), (c). For Figure.~\ref{fig:cE_emission} (b) the relvant Lagrangian interaction at the tidal vertex is  $-2\sigma_{ab}(\partial^a\partial^i\phi)(\partial_i\partial^b\phi)$,  while for Figure.~\ref{fig:cE_emission} (c) at the tidal vertex one has: $\int \td t (\partial^i\partial^j\phi)(\partial_i\partial_j\phi)$.
The corresponding effective Lagragians are
\begin{align}
\label{eq:c_Efig2b}
&i\mathcal{L}_{\text{2b}}=-\frac{c_E^{1}m_2^2}{32m^5_{\text{pl}}}\sigma_{ab}\left[\int_{\mathbf{q}}\frac{e^{i\mathbf{q}\cdot\mathbf{r}}}{q^2}q^a q_i\int_{\mathbf{k}}\frac{e^{i\mathbf{k}\cdot\mathbf{r}}}{k^2}k_i k^b
\right]\notag\\\notag    &~~~~~~~~~~~~~~~~~~~~~
+(m_1\leftrightarrow m_2)\\
    &~~~~~~~~~=-\frac{6iG^2c_E^1m_2^2}{r^6}\hat n^a\hat n^b\frac{\sigma_{ab}}{m_{\text{pl}}}+(m_1\leftrightarrow m_2)\,,
\end{align}
\begin{align}
\label{eq:c_Efig2c}
&i\mathcal{L}_{\text{2c}}=\frac{c_E^{1}m_2^2}{32m^5_{\text{pl}}}\sigma_{ab}\left[
\int_{\mathbf{q}}
\frac{e^{i\mathbf{q}\cdot\mathbf{r}}}{q^2}q^i q^j\int_{\mathbf{k}}
\frac{e^{i\mathbf{k}\cdot\mathbf{r}}}{k^4}
k_i k_j k^ak^b \right]\notag\\\notag
    &~~~~~~~~~~~~~~~~~~~~~+(m_1\leftrightarrow m_2)\\
    &~~~~~~~~~=-\frac{12iG^2c_E^1m_2^2}{r^6}\hat n^a\hat n^b\frac{\sigma_{ab}}{m_{\text{pl}}}+(m_1\leftrightarrow m_2),
\end{align}

Addition of eqs.~(\ref{eq:c_Efig2b}) and Eq.~(\ref{eq:c_Efig2c}), to Eq.~(\ref{eq:c_Efig2_a}) gives the full radiative effective action linear in $c_E$ in Eq.~(\ref{eq:S_eff_rad_cE}).

In addition, bulk vertices in Eq.~(\ref{eq:LWRt3phi3sigma}) are supposed to give rise to the counterterm as $\notag\int \td t\, \sigma_{ab}(\partial^a\partial^i\phi)(\partial_i\partial^b\phi)$. 
One can verify that the corresponding counterterms of Eq.~(\ref{eq:LWRt3phi3sigma}) are just $-2c_E^{pp}\int \td t\, \sigma_{ab}(\partial^a\partial^i\phi)(\partial_i\partial^b\phi)$, where $c_E^{pp}$ is the Wilson coefficient in front of the counterterm $\int \td t (\partial^i\partial^j\phi)(\partial_i\partial_j\phi)$ for each theory. 

\section{Redundancy of $R_{abcd} R^{abcd}$} \label{sec:R2}
As a last example, let us review our approach beyond the leading order contribution in a case that should give rise
to no physical change. 
As mentioned, a higher-curvature term is redundant if it can be removed either by a field redefinition, such as the $RK$ term, or by topological identities. While we have already analysed how to eliminate non-physical contributions arising from the former class, it is instructive to examine our approach with higher-curvature operators built from topological invariants. As a representative example, we consider the operator $R_{abcd} R^{abcd}$, which vanishes in vacuum in four dimensions due to the Gauss-Bonnet invariant. We expect that the variation of $R_{abcd} R^{abcd}$ can be removed by the ``trivial'' worldline operators completely. For simplicity, we work in the static NRG gauge.

Applying our method for this case, it is easy to show that its leading order contribution lies at 3PN, and gives the corresponding term
\begin{equation}
    \int \td^4 x (\partial_i\partial_j\phi)(\partial^i\partial^j\phi),
\end{equation}
whose variation gives
    $\square^2\phi$,
which does not belong to the two classes of source that gives divergent solutions, since it is linear in $\phi$. In fact, at the leading order, the $\alpha R^{abcd}R_{abcd}$ vertex gives the solution of a massive graviton of mass $1/(2\sqrt{\alpha})$ with the effective potential~\cite{Stelle1977}
\begin{equation}
    V_{3PN}\sim \frac{e^{-r/\sqrt{4\alpha}}}{r},
\end{equation}
which decays exponentially for $\alpha\xrightarrow[]{}0^+$. 
Or equivalently, at 3PN order, it gives a ``contact term'' as the solution $\delta\phi=-16\pi Gm\delta(x)$. Given this, we have to proceed to the next-to-leading order to obtain counterterms and show that our method is effective.
We denote $\phi$ and $\gamma_{ij}$ as follows
\begin{align}\label{eq:PN_NRG_metric}
    &\phi=\phi^{(1)}+\phi^{(2)}+\cdots\\
    &\gamma_{ij}=\delta_{ij}+\sigma_{ij}=\delta_{ij}(1+\sigma)=\delta_{ij}(1+\sigma^{(1)}+\sigma^{(2)}+\cdots),
\end{align}
where the superscript ``$(n)$'' represents the $r^{-n}$ order of the field. $\sigma_{ij}$ reduces to $\sigma\delta_{ij}$ due to spherical symmetry.
As shown in Appendix.~\ref{app:GR_pp}, all the $\phi^{(2k)}$ and $\sigma^{(2k-1)}$ ($k\in \mathbb{N}^+$) vanish.
As a result, at 4PN order, the on-shell equation of motion of $\delta\phi^{4PN}$ only evolves $\sigma^{(2)}$ and $(\phi^{(1)})^2$
\begin{align}
    &\square\delta\phi^{4PN}+\left((\sigma^{(2)},(\phi^{(1)})^2 \text{ by varying }R_{abcd} R^{abcd}\right)=0.
\end{align}
The $\sigma\phi$ term of $R_{abcd} R^{abcd}$ is
\begin{equation}
    -4\partial^b\partial^a\phi\partial_b\partial_a\text{tr}(\sigma)+8\partial^b\partial^a\phi\partial_c\partial_a\sigma^c_b-4\partial^b\partial^a\phi\partial^2\sigma_{ab},
\end{equation}
which comes into the equation of motion as
\begin{equation}\label{eq:EOM_sigma}
4\partial^b\partial^a\square\sigma_{ab}-4\square^2\text{tr}(\sigma).
\end{equation}
While the $\phi^3$ cubic term of $\sqrt{-g}R^{abcd}R_{abcd}$ is
\begin{align}
    &-16\partial_a\phi\partial^a\phi\square\phi+32\partial^a\phi\partial^b\phi\partial_a\partial_b\phi+8\phi(\square\phi)^2\notag\\
    &~~~+16\phi\partial_b\partial_a\phi\partial^b\partial^a\phi
\end{align}
with the equation given by,
\begin{equation}\label{eq:EOM_phi}
    88(\square\phi)^2-16\partial_a\partial_b\phi\partial^a\partial^b\phi+80\partial^i\phi\partial\square\phi+48\phi\square^2\phi.
\end{equation}
After implementing $\sigma_{ij}=\sigma\delta_{ij}$, the leading order equation of motion $\sigma$ satisfies
\begin{equation}\label{eq:1PN_sigma}
    \square\sigma^{(2)}=-(\partial_i\phi^{(1)})^2,
\end{equation}
which is obtained by varying the EH action.
Then, using this relation, Eq.~(\ref{eq:EOM_sigma}) reduces to
\begin{align}
    &8\square\left[(\partial_i\phi^{(1)})(\partial^i\phi^{(1)})\right]\notag\\
    &\xrightarrow[]{}16(\partial_i\partial_j\phi^{(1)})(\partial^i\partial^j\phi^{(1)})+16\partial_i\square\phi^{(1)}\partial^i\phi^{(1)}.
\end{align}
Together with Eq.~(\ref{eq:EOM_phi}), they contribute to the equation of motion of $\delta\phi^{4PN}$ as
\begin{equation}   40(\square\phi^{(1)})^2+48\square(\phi^{(1)}\square\phi^{(1)}),
\end{equation}
where both terms are of the type give in equation~(\ref{eq:ct1}) and can thus be removed by adding worldline operators $\int \td\tau \delta(x(\tau))\phi$. Hence, we show that for the $R_{abcd} R^{abcd}$ operator, which is supposed to vanish, our prescription is effective at 4PN order. Last, this result complements the one discused in~\cite{Bernard:2025dyh} by indicating this operator would not yield a non-trivial
counterterm.

\end{document}